\documentclass[pra,floatfix,footinbib,reprint,superscriptaddress]{revtex4-1}

\usepackage{iopams,setstack,amsopn} 
\expandafter\let\csname equation*\endcsname\relax
\expandafter\let\csname endequation*\endcsname\relax
\usepackage{amsmath, amssymb, amsthm, amsfonts}
\usepackage{graphicx}
\usepackage{dcolumn}
\usepackage{times}
\usepackage{dsfont}
\usepackage{tabularx}
\usepackage{subfigure}
\usepackage[e]{esvect}
\usepackage{tikz}
\usepackage{tablefootnote}
\usetikzlibrary{matrix}
\usepackage[pdftex, colorlinks, allcolors=blue]{hyperref}
\DeclareSymbolFont{sfletters}{OML}{cmbrm}{m}{it}

\DeclareMathOperator{\Tr}{Tr}

\DeclareMathOperator{\sinc}{sinc}
\DeclareMathOperator{\tanc}{tanc}

\newcommand{\siswap}{\sqrt{i\mathrm{SWAP}}}
\newcommand{\sbswap}{\sqrt{b\mathrm{SWAP}}}
\newcommand{\sphiswap}{\sqrt{\phi\mathrm{SWAP}}}

\newcommand{\oket}[1]{|\overline{#1}\rangle}
\newcommand{\obra}[1]{\langle\overline{#1}|}

\newcolumntype{Y}{>{\centering\arraybackslash}X}

\begin{document}

\title{Fast high-fidelity gates for galvanically-coupled fluxonium qubits using strong flux modulation}

\author{D.~K.~Weiss}
\email{dkweiss@u.northwestern.edu}
\address{Department of Physics and Astronomy, 
Northwestern University, Evanston, Illinois 60208, USA}
\author{Helin Zhang}
\affiliation{James Franck Institute, University of Chicago, Chicago, Illinois 60637, USA}
\affiliation{Department of Physics, University of Chicago, Chicago, Illinois 60637, USA}
\author{Chunyang Ding}
\affiliation{James Franck Institute, University of Chicago, Chicago, Illinois 60637, USA}
\affiliation{Department of Physics, University of Chicago, Chicago, Illinois 60637, USA}
\author{Yuwei Ma}
\affiliation{James Franck Institute, University of Chicago, Chicago, Illinois 60637, USA}
\affiliation{Department of Physics, University of Chicago, Chicago, Illinois 60637, USA}
\author{David I. Schuster}
\affiliation{James Franck Institute, University of Chicago, Chicago, Illinois 60637, USA}
\affiliation{Department of Physics, University of Chicago, Chicago, Illinois 60637, USA}
\affiliation{Pritzker School of Molecular Engineering, University of Chicago, Chicago, Illinois 60637, USA}
\author{Jens Koch}
\address{Department of Physics and Astronomy, 
Northwestern University, Evanston, Illinois 60208, USA}
\date{\today}

\begin{abstract}

Long coherence times, large anharmonicity and robust charge-noise insensitivity render fluxonium qubits an interesting alternative to transmons. Recent experiments have demonstrated record coherence times for low-frequency fluxonia. Here, we propose a galvanic-coupling scheme with flux-tunable $\textit{XX}$ coupling. 
To implement a high-fidelity entangling $\siswap$ gate, we modulate the strength of this coupling and devise variable-time identity gates to synchronize required single-qubit operations. Both types of gates are implemented using strong ac flux drives, lasting for only a few drive periods. We employ a theoretical framework capable of capturing qubit dynamics beyond the rotating-wave approximation (RWA) as required for such strong drives.
We predict an open-system fidelity of $F>0.999$ for the $\siswap$ gate under realistic conditions. 
\end{abstract}

\maketitle

\section{Introduction}

A major challenge in the field of quantum computing is to break free from the imperfections characteristic of the noisy intermediate-scale quantum (NISQ) \cite{NISQ} era. For superconducting qubits, this will specifically require further improvements of two-qubit gate performance beyond the current state-of-the-art, with errors on the order of $10^{-3}-10^{-4}$ \cite{Negirneac2021, Dogan2022, Foxen2020}. To reach even lower two-qubit gate infidelities, it is worth re-examining the framework routinely used for developing the pulse trains which generate the gates of interest. Most commonly, this framework is intimately linked to the use of the rotating-wave approximation (RWA). This approximation is highly convenient as it can help remove fast time dependence from the Hamiltonian, yields an intuitive picture of the dynamics, and makes calculations particularly simple \cite{Blais2021}. However, the range of validity of the RWA is limited and reliance on it constrains the parameter space explorable for maximizing gate fidelities. For low-frequency qubits such as heavy fluxonium \cite{Earnest2018, Lin2018, Zhang2020}, this is particularly unfortunate as higher gate fidelities can indeed be achieved for parameter choices outside the reach of the RWA. 

Here, we employ a theoretical framework based on the Magnus expansion \cite{Magnus, Blanes2009, Wilcox, Zeuch2020}, supplemented by full numerics, for executing high-fidelity gates in the regime of strong driving where drive amplitudes approach or even exceed the qubit frequency \cite{Zhang2020, Campbell2020, Yang2017, Wu2007}. 
Motivated by the hundreds of microseconds \cite{Zhang2020, Nguyen2019} to millisecond \cite{Somoroff2021} coherence times recently observed in low-frequency fluxonium qubits, we apply this analysis to a coupled system of such fluxonia. Two-qubit gates on capacitively-coupled fluxonia have recently been reported \cite{Ficheux2021, Xiong2021arbitrary, bao2021fluxonium, Dogan2022, Moskalenko2022}, with infidelities on the order of $10^{-2}-10^{-3}$. These experimental realizations have been accompanied by a flurry of theoretical attention \cite{Chen2021, nesterov2021, moskalenko2021, nesterov2022, Cai2021, Nesterov2018, Nguyen2022}. Fluxonium qubits in these capacitive-coupling architectures have frequencies on the order of 500 MHz - 1 GHz, generally thought to be the ideal frequency range for executing high-fidelity gates in these systems (considering gate schemes involving population transfer only in the qubit subspace) \cite{nesterov2021, Chen2021, nesterov2022}. However, fluxonia with qubit frequencies less than 200 MHz have consistently achieved the longest coherence times \cite{Somoroff2021, Zhang2020}, likely because low-frequency operation mitigates dielectric loss \cite{Nguyen2019, Zhang2020, Somoroff2021}. To obtain fast, high-fidelity entangling gates between such low-frequency fluxonium qubits, we revisit inductive-coupling schemes previously proposed \cite{FQCT1, FQCT2, FQCT3} and experimentally implemented \cite{FQCE1, FQCE2} for flux qubits.

We consider fluxonium qubits linked galvanically via a flux-tunable coupler. To avoid directly exciting the coupler degrees of freedom, the interaction between the qubits and the coupler is chosen to be dispersive. This allows for an effective description in which the coupler is eliminated, but mediates a tunable $\textit{XX}$ interaction. We show that the strength of this effective $\textit{XX}$ coupling changes sign as a function of coupler flux and thus passes through zero. In addition, we find that the parasitic $\textit{ZZ}$ interaction strength is suppressed, which is a general feature of coupled systems of low-frequency fluxonium qubits \cite{Ficheux2021}. 

We describe how to execute two-qubit gates via sinusoidal modulation of the coupler flux for a duration as short as a few drive periods. Based on our analysis outside the RWA regime, we find that the implemented entangling operations generally differ from named gates by relative phases. We compensate for these phase factors using single-qubit $\textit{Z}$ rotations to obtain a high-fidelity $\siswap$ gate.

Ordinarily, single-qubit gates are designed in the convenient regime where RWA applies \cite{Blais2021}. In this scenario, the switch into a  frame co-rotating with the drive renders operations simple  rotations about fixed axes in the Bloch-sphere picture. Identity gates are obtained by idling and single-qubit $\textit{Z}$ rotations are obtained either e.g. ``virtually" by modifying the phase of the drive field in software \cite{McKay2017, Chen2021phase} or ``physically" by detuning the qubit frequency from that of the drive \cite{Lucero2010}. The situation is reversed for systems of heavy fluxonium qubits \cite{Earnest2018, Lin2018} where drive strengths exceeding the RWA-range are employed -- motivating the use of the laboratory frame for qubit operations \cite{Zhang2020, Campbell2020}. In this frame, qubits acquire dynamical phases in the absence of control pulses; in  other words, idling yields $\textit{Z}$ rotations of each qubit \cite{Zhang2020, Campbell2020}. To synchronize gates in multi-qubit systems, identity gates of variable time duration must be devised. We show that identity gates can again be implemented using sinusoidal modulation of the qubit fluxes for only a single drive period, resulting in ultra-fast $I$ gates  (when compared to the single-qubit Larmor period). Combining the identity gates with single-qubit $\textit{Z}$ rotations assists in achieving high-fidelity entangling gates.

This paper is organized as follows. We introduce the coupling scheme in Sec.~\ref{sec:fullcircuit}, and derive the full-circuit Hamiltonian and its effective counterpart governing the low-energy physics. We describe the implementation of single-qubit gates in Sec.~\ref{sec:single_q_gates}  focusing specifically on the needed identity gates. In Sec.~\ref{sec:two_q_gate} we detail our scheme for performing high-fidelity two-qubit gates, including the necessary single-qubit $\textit{Z}$ rotations and identity gates. In Sec.~\ref{sec:conc} we summarize our results and provide an outlook on future work.

\section{Galvanically-coupled heavy-fluxonium qubits}
\label{sec:fullcircuit}

\begin{figure}
  \includegraphics[width=\columnwidth]{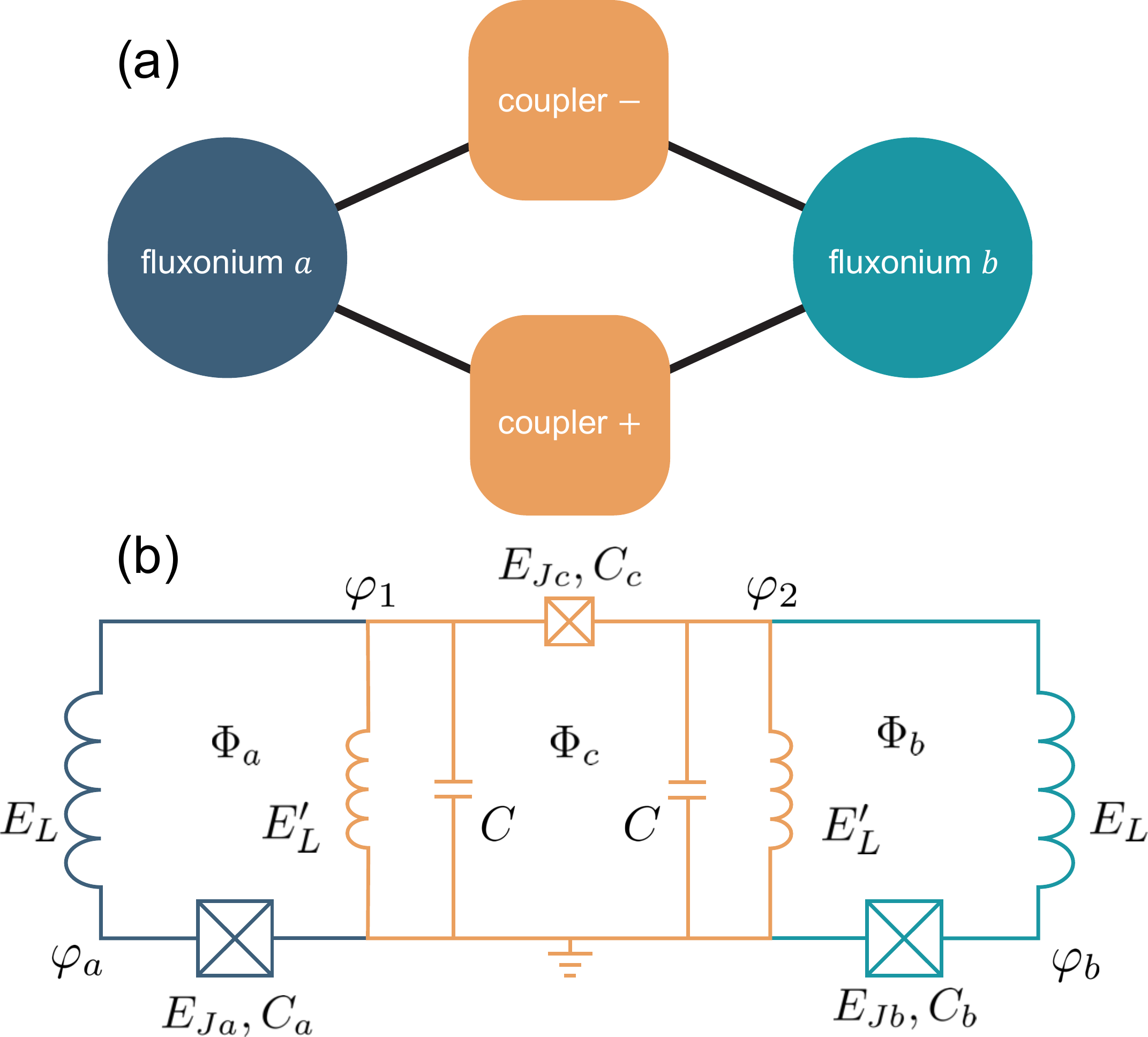}
  \caption{\label{fig:circuit} Galvanically-coupled heavy fluxonia. (a) Schematic of the qubit-coupler interaction. The fluxonium qubits $a, b$ are each coupled to a harmonic coupler mode $\theta_{+}$ and a flux-tunable coupler mode $\theta_{-}$, but do not interact directly. (b) Circuit diagram of the device. Qubit $a$ (dark blue) and qubit $b$ (light blue) are galvanically linked to the two coupler modes $\theta_{\pm}$ (orange). Each loop can be threaded by an external flux 
  $\Phi_{\mu}$. }
\end{figure}

We consider galvanically linking two fluxonium qubits via a flux-tunable coupler.
We bias the heavy-fluxonium qubits at their half-flux sweet spots, as such fluxonia are linearly insensitive to flux noise and thus 
have achieved record coherence times \cite{Somoroff2021, Zhang2020, Nguyen2019}.
The qubit states are delocalized over two neighboring wells of the fluxonium potential, with the qubit frequency set by the tunnel splitting \cite{Koch2009, Manucharyan2009}. Single-qubit gates can be achieved by tuning the external flux away from the sweet spot, activating a transverse $\textit{X}$ interaction \cite{Zhang2020}. A galvanic-coupling scheme helps yield strong coupling strengths, as those are quantified by phase matrix elements rather than charge matrix elements (as would be the case for capacitive coupling). Phase matrix elements are not suppressed by the qubit frequency \cite{Nesterov2018}, making a galvanic-coupling architecture attractive for achieving entangling gates on low-frequency qubits.
Moreover, for fluxonium qubits whose inductance is kinetic \cite{Manucharyan2009} rather than geometric \cite{Peruzzo2021}, a galvanic connection can generally yield stronger coupling strengths than those achieved through mutual inductance alone \cite{Peruzzo2021}. To make the coupling strength tunable, we generalize the so-called ``fluxonium molecule" circuit introduced in Ref.~\cite{Kou2017} by inserting 
a coupler Josephson junction as shown in Fig.~\ref{fig:circuit}. 
The circuit Hamiltonian is $H=H_{0}+V$, where
\begin{align}
\nonumber 
H_{0} &=\sum_{\mu=a,b}[4E_{C\mu}n_{\mu}^2+\frac{1}{2}E_{L}\varphi_{\mu}^2-E_{J\mu}\cos(\varphi_{\mu}+\pi)] \\
\label{eq:H0}
&\quad+ 4E_{C-}n_{-}^2+\frac{1}{2}E_{Lc}\theta_{-}^2-E_{Jc}\cos(\theta_{-}+\phi_{c}) \\ \nonumber 
&\quad+ 4E_{C+}n_{+}^2+\frac{1}{2}E_{Lc}\theta_{+}^2, \\
\label{eq:V}
V&=- \frac{E_{L}}{2}[\varphi_{a}(\theta_{+}+ \theta_{-})
+\varphi_b(\theta_{+}- \theta_{-})] \\ \nonumber 
&\quad+\sum_{\mu=a,b}\frac{E_{L}}{2}{\delta\phi_{\mu}}[-2\varphi_{\mu}+\theta_{+}+(-1)^{\mu}\theta_{-}],
\end{align}
where $\mu$ obeys the correspondence $a\to0, b\to1$ when appearing in an exponent. See Appendix \ref{appendix:fullcircuit} for details on the full derivation of the Hamiltonian $H$. We have defined the inductive energy of the coupler $E_{Lc}=\frac{1}{2}(E_{L}+E_{L}')$, and the charging energies $E_{C\mu}=e^2/2C_{\mu},\;\mu=a,b, E_{C+}=e^2/(2[C/2]), E_{C-}=e^2/(2[C_{c}+C/2])$, where the remaining circuit parameters can be read off from Fig.~\ref{fig:circuit}. The node variables $\varphi_{a}, \varphi_{b}$ are the qubit variables, while the coupler variables are defined as $\theta_{\pm}=\varphi_{1}\pm\varphi_{2}$. We have isolated the qubit-flux shifts away from the sweet spot $\delta\phi_{\mu}=\phi_{\mu}-\pi$, where $\phi_{\mu}=2\pi\Phi_{\mu}/\Phi_{0}$ is the reduced external flux, $\Phi_{\mu}$ is the external flux in the corresponding loop and $\Phi_{0}=h/2e$ is the superconducting flux quantum. 

The Hamiltonian $H$ is composed of two fluxonium qubits and two coupler degrees of freedom, where the qubits interact with the coupler via terms in $V$. The coupler $\theta_{+}$ degree of freedom is harmonic, while the coupler $\theta_{-}$ degree of freedom is fluxonium like and thus tunable by external flux.  
It is important to note that there is no term in the Hamiltonian directly coupling the qubits, thus the qubit-qubit interaction is entirely mediated by the coupler. 
Here we assume symmetric qubit inductors, coupler inductors, and stray coupler capacitances, respectively, see Fig.~\ref{fig:circuit}. In this case, there are no terms in the Hamiltonian that directly couple the coupler degrees of freedom. This assumption is relaxed in Appendix \ref{appendix:fullcircuit} where we derive the Hamiltonian in the presence of disorder in circuit parameters.

The coupler should have the following two desired properties.
The first is it must allow for the execution of high-fidelity two-qubit gates. A straightforward way to achieve this goal is to ensure that the interaction of the qubits with the coupler degrees of freedom is dispersive, allowing for an effective description in terms of two coupled qubits. The second requirement is that the two-qubit coupling strength should be sufficiently flux dependent, allowing for tuning from zero to values that allow for fast gates compared with the coherence times $T_{1}, T_{2}$ of each qubit. In the next section, we derive the effective Hamiltonian of the system assuming that the qubit-coupler interaction is dispersive. Following this, we discuss specific coupler parameter choices that satisfy the above requirements.

\subsection{Low-energy effective Hamiltonian}
\label{subsec:effectiveHam}

Near the half-flux sweet spots for each qubit, the qubit excitation energies are small compared with the energy needed to excite the coupler or higher-lying fluxonium states. These energy scales naturally define two subspaces: the low-energy subspace defined by the projector onto the computational states $P=\sum_{\ell,m=0,1}|\ell_{a},m_{b},0_{-},0_{+}\rangle\langle\ell_{a},m_{b},0_{-},0_{+}|$ and the high-energy subspace spanned by all other states.
We have defined the bare states $|\ell_{a},m_{b},n_{-},p_{+}\rangle$ that are eigenstates of $H_{0}$ with eigenenergies $E_{l}^{a}+E_{m}^{b}+E_{n}^{-}+p\omega_{+}$. The variables $\ell,m,n,p$ correspond to the number of excitations in the degrees of freedom corresponding to the variables $\varphi_{a}, \varphi_{b}, \theta_{-}, \theta_{+}$, respectively. 
\begin{figure}
    \centering
    \includegraphics[width=\columnwidth]{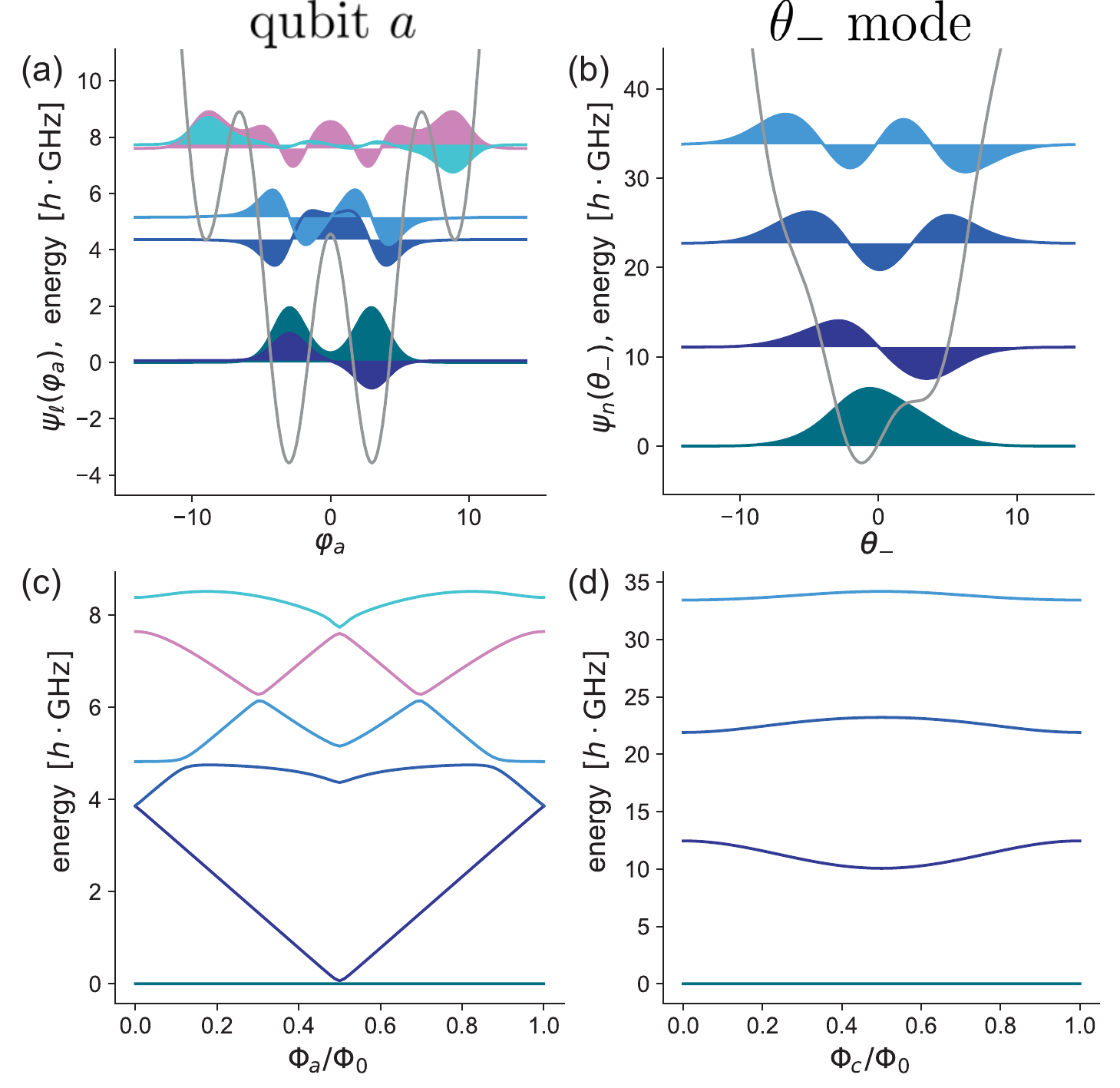}
    \caption{\label{fig:qubit_coupler_wf_energies} 
    Bare wave functions and energy spectra. (a) Bare wave functions and potential of qubit $a$ located at the half-flux sweet spot. The bare qubit transition frequency is $\omega_{a}/2\pi=62$ MHz, while the energy of the next excited state is 4.4 $h\cdot$GHz above the qubit ground-state energy. (b) Spectrum of qubit $a$ as a function of the qubit flux $\Phi_{a}$. The wave functions and spectra of qubit $b$ are qualitatively similar to those of qubit $a$. (c) Bare wave functions and potential of the coupler $\theta_{-}$ mode located at $\Phi_{c}/\Phi_{0}=0.27$. The energy of the first excited state is more than 10 $h\cdot$GHz above the coupler ground-state energy. (d) Spectrum of the coupler $\theta_{-}$ mode as a function of the coupler flux $\phi_{c}$. See Table~\ref{table:params} for device parameters.}
\end{figure}
\begin{table}
\caption{\label{table:params}Circuit parameters in $h\cdot$GHz used throughout this work. 
}
\centering
\begin{tabularx}{\linewidth}{*9{>{\centering\arraybackslash}X}}
\hline\hline
$E_{Ja}$ & $E_{Jb}$ & $E_{Ca}$ & $E_{Cb}$ & $E_{L}$
& $E_{Jc}$ & $E_{L}'$ & $E_{C-}$ & $E_{C+}$
\\
\hline
 4.6 & 5.5 & 0.9 & 0.9 & 0.21 & 3 & 2 & 14.3 & 100 \\
 \hline
 \hline
\end{tabularx}
\end{table}
The coupler $\theta_{-}$ mode is fluxonium like,
however operated in a different parameter regime than the heavy-fluxonium qubits. The bare wave functions and spectra of qubit $a$ and the coupler $\theta_{-}$ mode are shown in Fig.~\ref{fig:qubit_coupler_wf_energies}. We discuss below in detail the ideal parameter regime in which to operate the coupler modes. 

States in different subspaces are coupled by the perturbation $V$, which is small compared with the relevant energy separations. Thus, the interaction is dispersive and we can obtain an effective description of the low-energy physics via a Schrieffer-Wolff transformation \cite{Blais2021, SW, *cohentannoudji, *Winkler2003}. This is done by introducing a unitary $e^{-S}$ with anti-hermitian generator $S$ that decouples the high- and low-energy subspaces order-by-order. We find the effective Hamiltonian up to second order upon projecting onto the low-energy subspace
\begin{align}
\label{eq:effHam}
H_{\text{eff}} = -\sum_{\mu=a,b}\frac{\omega_{\mu}'}{2}\sigma_{z}^{\mu} + J\sigma_{x}^{a}\sigma_{x}^{b} -\sum_{\mu=a,b}\Omega_{\mu}\sigma_{x}^{\mu}.
\end{align}
This transformation is carried out in detail in Appendix \ref{appendix:SWT}.
The Hamiltonian $H_{\mathrm{eff}}$ describes two qubits with frequencies $\omega_{\mu}'=\omega_{\mu}+\chi_{\mu}$, where $\omega_{\mu}=E_{1}^{\mu}-E_{0}^{\mu}$ are the bare qubit frequencies (here and in the following we set $\hbar=1$) and $\chi_{\mu}$ are the Lamb shifts. The qubits are coupled via a transverse $\textit{XX}$ interaction with strength $J$ that is tunable with coupler flux $\phi_{c}$. There are additional single-qubit $\textit{X}$ terms with strength $\Omega_{\mu}$ that depend on the coupler flux $\phi_{c}$ as well as the qubit fluxes $\delta\phi_{\mu}$. We show below that both $J$ and $\Omega_{\mu}$ can be tuned through zero, yielding two decoupled qubits. 
The coefficient $\Omega_{\mu}$ is defined as
\begin{align}
\label{eq:Omega}
\Omega_{\mu}=E_{L}\langle 0_{\mu}|\varphi_{\mu}|1_{\mu}\rangle\left[{\delta\phi_{\mu}}+(-1)^{\mu}\frac{\langle 0_{-}|\theta_{-}|0_{-}\rangle}{2}\right],
\end{align}
and arises at first-order in perturbation theory from two contributions. The first term on the right-hand side is due to a qubit-flux offset from the sweet spot $\delta{\phi_{\mu}}$, while the second is from the coupling between the qubits and the coupler $\theta_{-}$ mode. The matrix element $\langle0_{-}|\theta_{-}|0_{-}\rangle$ is $\phi_{c}$ dependent and generally nonzero due to the absence of selection rules for a fluxonium biased away from a sweet spot \cite{Zhu2013a, Zhu2013b}. Interpreting this second term as an effective flux shift away from the sweet spot for each qubit, we cancel this shift 
by setting
\begin{align}
\label{eq:deltaphistat}
\delta{\phi_{\mu}}&=(-1)^{\mu+1}\frac{\langle0_{-}|\theta_{-}|0_{-}\rangle}{2}.
\end{align}
\begin{figure}
    \centering
    \includegraphics[width=0.7\columnwidth]{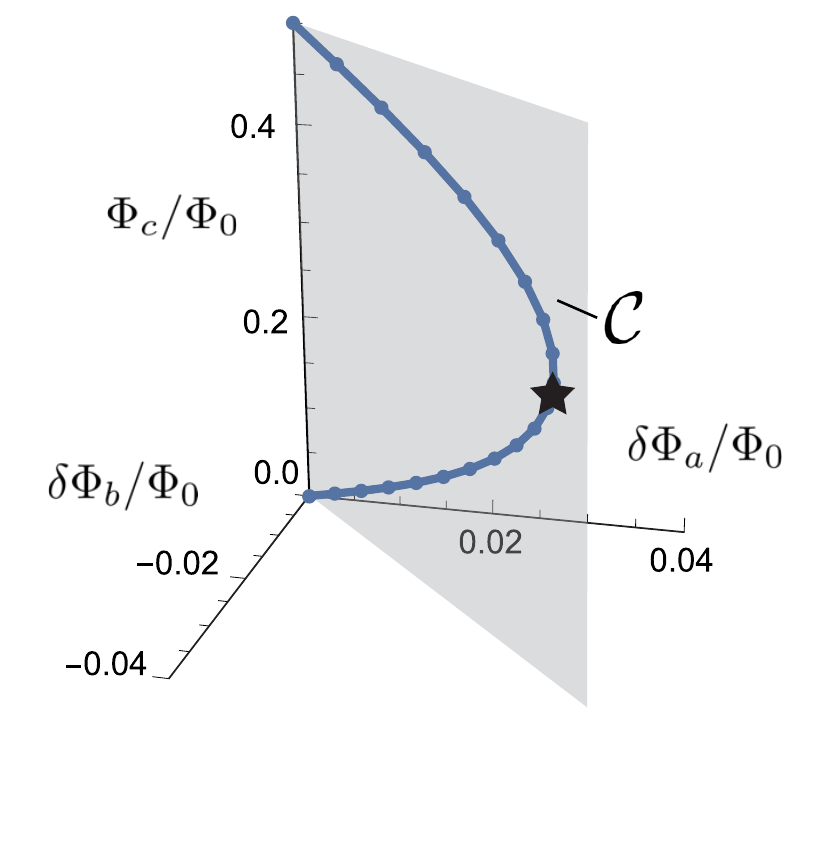}
    \caption{\label{fig:flux_contour}Sweet-spot contour. As the coupler flux is tuned, the qubit fluxes must be simultaneously adjusted according to $\delta\Phi_{\mu}/\Phi_{0}=(-1)^{\mu+1}\frac{\langle 0_{-}|\theta_{-}|0_{-}\rangle}{4\pi}$ in order to keep the qubits at their sweet spots $\Omega_{\mu}=0$. The qubit fluxes 
    are measured as deviations away from the bare sweet-spot locations. The sweet-spot contour $\mathcal{C}$ lies in the semi-transparent gray plane, which bisects the angle between the two qubit-flux axes. The off position where $\Omega_{\mu}=J=0$ is marked by a black star.
    }
\end{figure}
Thus, we obtain the coupler-flux dependent ``sweet-spot contour" shown in Fig.~\ref{fig:flux_contour}. 
It is important to note that this phenomenon is independent of effects due to geometric flux crosstalk and arises instead directly from coupling terms in $V$. 

The two-qubit interaction arises at second-order in perturbation theory, with strength $J=J_{-}-J_{+}$. The coefficient $J_{-}$ ($J_{+}$) is due to interaction of the qubit with the coupler $\theta_{-}$ ($\theta_{+}$) mode. The strength of the interaction $J_{-}$ is tunable due to the dependence of the matrix elements $\langle 0_{-} | \theta_{-}| n_{-}\rangle$ and energies $E_{n}^{-}$ on coupler flux, see Appendix~\ref{appendix:SWT} for details. 
The coefficient $J_{+}$ is static, thus the two-qubit interaction is eliminated by tuning $J_{-}$ to equal $J_{+}$ in magnitude \cite{Mundada2019}. We can understand the $\textit{XX}$ nature of the two-qubit coupling by considering the terms appearing in $V$ that mediate the interaction between different subsystems. At the qubit sweet spots and considering only the computational states, the operators $\varphi_{\mu}$ are off-diagonal and therefore proportional to $\sigma_{x}^{\mu}$ with proper choice of phases. Thus, the effective two-qubit interaction consists of a virtual second-order process whereby an excitation is exchanged between the two qubits, or both qubits are co-excited or co-de-excited. 

In what follows, we always operate from the dc flux bias point on the sweet-spot contour $\Omega_{\mu}=0$ where the two-qubit coupling is turned off $J=0$, see Fig.~\ref{fig:flux_contour}. We do this to keep both qubits at their respective sweet spots and to prevent any unwanted parasitic entanglement between the qubits. We refer to this configuration of dc fluxes as the ``off position." Both single- and two-qubit gates are performed by ac flux excursions about this point. Note that the value of the coupler flux $\phi_{c}$ at the off position is generally parameter dependent. 

\subsection{Parameter regime of the coupler}

To obtain an effective description in terms of two-coupled qubits, parameter choices for the coupler should support a dispersive interaction. Additionally, we require that the coupling strength be sufficiently flux dependent, allowing both for the execution of fast gates and for the interaction to be efficiently turned off. We quantify the dispersiveness of the interaction by calculating the Lamb shift $\chi_{a}$ (we obtain similar results in the following utilizing instead $\chi_{b}$).
Considering the requirements on flux dependence, we calculate the slope of the coupling strength $J_{-}$ with respect to $\Phi_{c}$ at the off position. We target parameters such that $|\partial_{\Phi_{c}}J_{-}|/h\approx 100$ MHz/$\Phi_{0}$ to achieve MHz level coupling strengths 
(implying fast gates compared with $T_{1}, T_{2}$) for small ac flux excursions $\Phi_{c}\lesssim 0.03\, \Phi_{0}$ where a linear relationship between the coupling strength $J$ and flux $\Phi_{c}$ is expected to be valid. This value of the slope also ensures that the device remains insensitive to typical $1/f$ flux noise amplitudes $A_{\Phi}\approx 1 \mu\Phi_{0}$ \cite{Hutchings2017}. 

\begin{figure}
    \centering
    \includegraphics[width=\columnwidth]{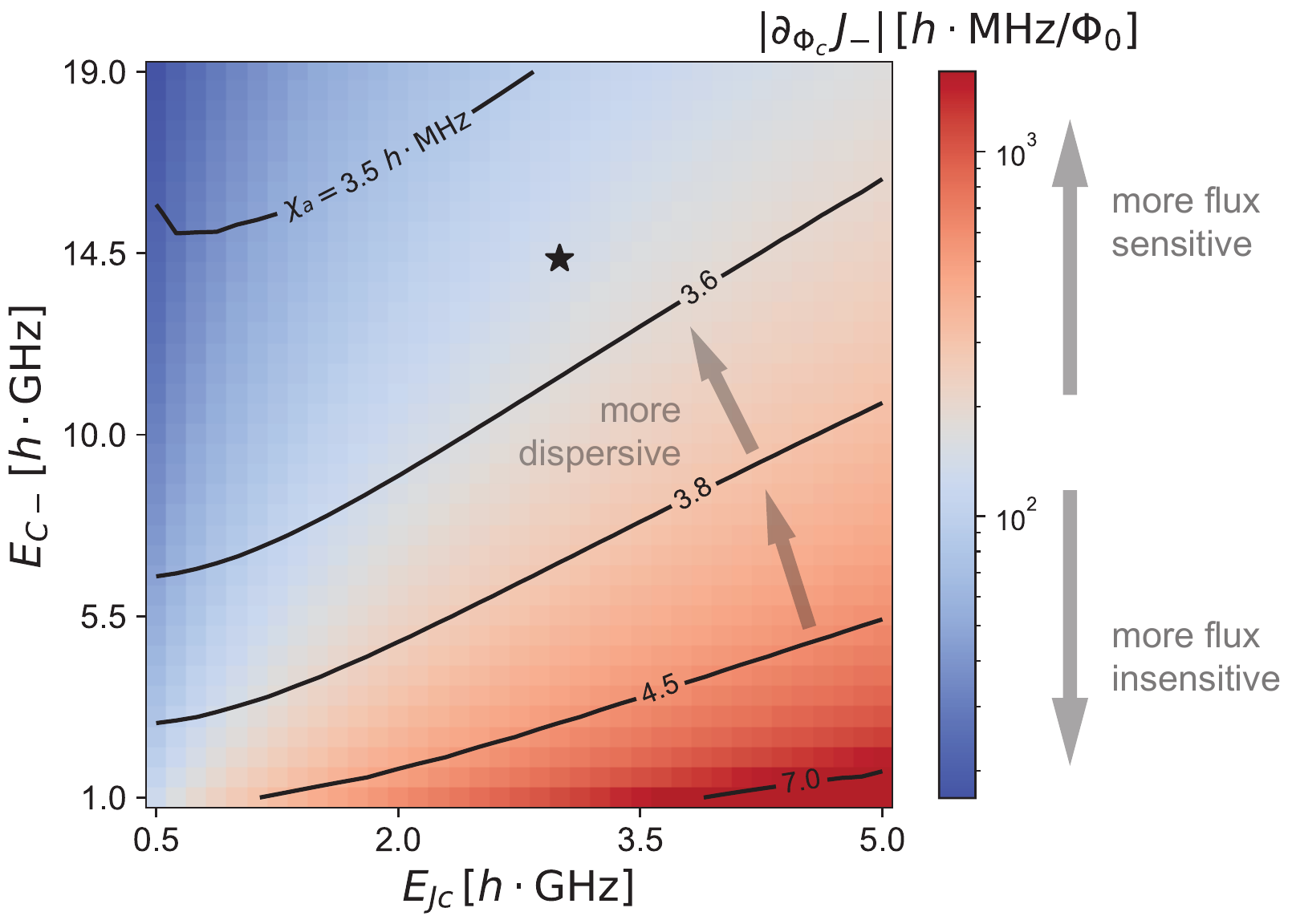}
    \caption{\label{fig:deriv_J} Flux sensitivity and dispersiveness as a function of the coupler parameters $E_{Jc}$ and $E_{C-}$.  Coloring indicates $|\partial_{\Phi_{c}}J_{-}|$, i.e., the linear sensitivity of the coupling strength with respect to flux. 
    Contour lines quantify the dispersiveness of the qubit-coupler interaction via the Lamb shift $\chi_{a}$. Dispersive interaction and suitable flux sensitivity are achieved in the parameter regime
    $E_{C-}>E_{Jc} \gtrsim E_{Lc}$. The star marks the chosen parameters, Table~\ref{table:params}.
    }
\end{figure}

We sweep over $E_{Jc}$ and $E_{C-}$ and calculate $|\partial_{\Phi_{c}}J_{-}|$ as well as $\chi_{a}$ at the off position, see Fig.~\ref{fig:deriv_J}. We fix $E_{L}'/h=2$ GHz ($E_{Lc}/h=1.1$ GHz), however we obtain similar results when considering instead larger or smaller values of $E_{L}'$. It is worth emphasizing that the off position is parameter dependent, thus we reposition the dc fluxes appropriately for each parameter set.
For relatively large $E_{Jc}$ and small $E_{C-}$, the lowest-lying states at intermediate flux values localize in minima of the cosine potential. The off position is then generally near the sweet spot, where the vanishing energy difference between the states $|0_{-}\rangle,|1_{-}\rangle$, as well as the rapid increase in the value of the matrix element $\langle 0_{-}|\theta_{-}|1_{-}\rangle$ enable flux tunability of $J_{-}$. These factors in turn imply extreme sensitivity to flux $|\partial_{\Phi_{c}}J_{-}|/h\gg \,100 \text{ MHz}/\Phi_{0}$ as well as a breakdown of the dispersive regime. For relatively small $E_{Jc}$ and large $E_{C-}$, flux tunability is lost as the spectrum is nearly harmonic. For decreasing $E_{Jc}$ and $E_{C-}$, excitation energies are suppressed leading to a breakdown of the dispersive interaction. The parameter regime that supports both a dispersive interaction and ``Goldilocks" flux dependence is thus $E_{C-}>E_{Jc}\gtrsim E_{Lc}$. 
The parameters $E_{C+}, E_{Lc}$ that define the coupler $\theta_{+}$ mode are implied by the parameter choices for the coupler $\theta_{-}$ mode, with the restriction $E_{C+}>E_{C-}$ due to the finite junction capacitance. The parameters used in the remainder of this work are given in Table~\ref{table:params}.

\subsection{Numerical results}
\begin{figure}
    \centering
    \includegraphics[width=\columnwidth]{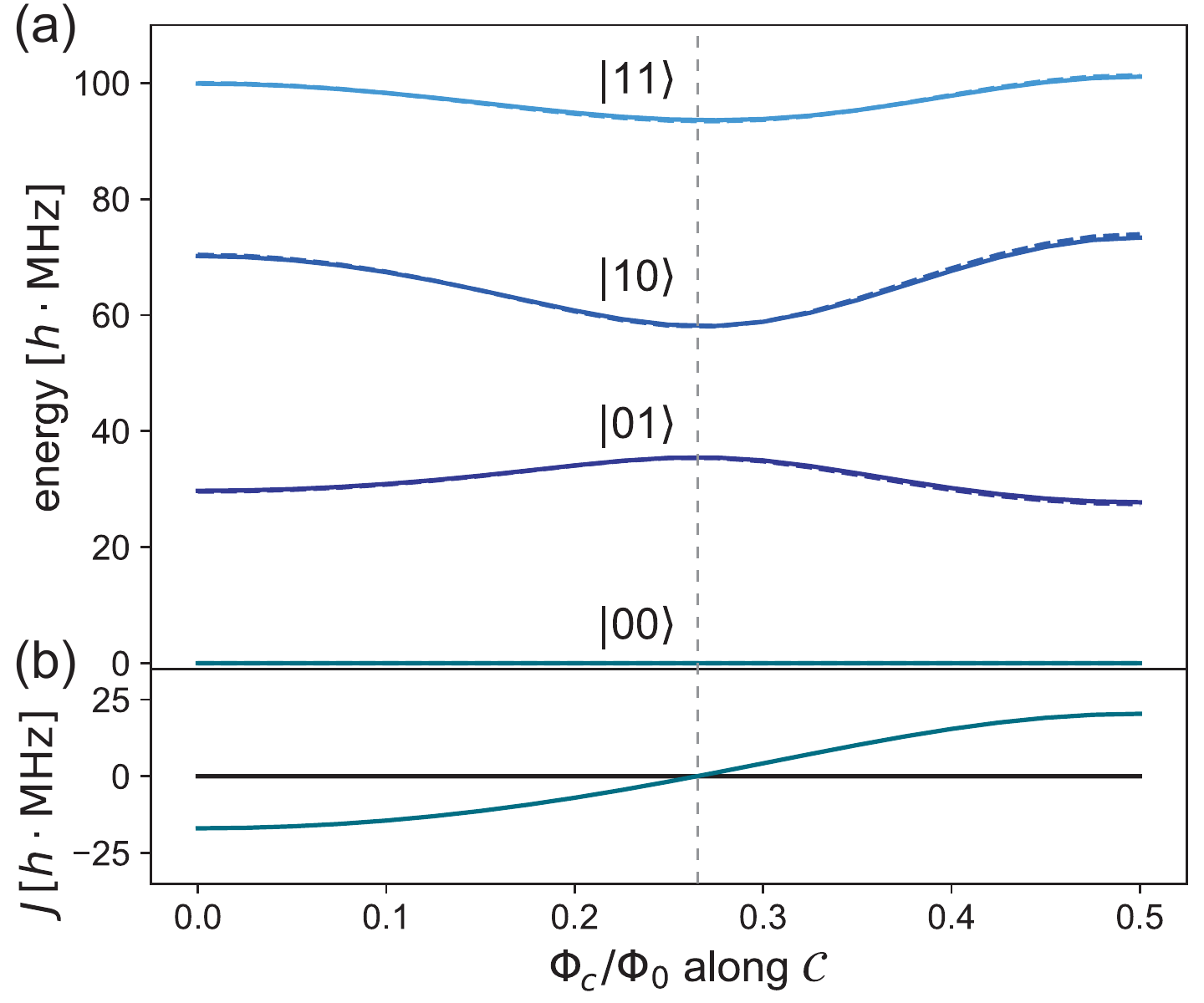}
    \caption{\label{fig:spectrum} (a) Low-energy spectrum of the coupled system. As the coupler flux is tuned, the qubit fluxes are adjusted to remain on the sweet-spot contour $\mathcal{C}$, ensuring $\Omega_{\mu}=0$. The full lines correspond to the exact spectrum calculated from the full model $H$, while the dashed lines correspond to the spectrum as calculated from the effective Hamiltonian $H_{\text{eff}}$. Eigenenergies are labeled according to the bare state $|ij\rangle\equiv|i_{a},j_{b},0_{-},0_{+}\rangle$ with the largest overlap with the corresponding dressed state when the coupler is in the off state (at $\phi_{c}\approx0.27\cdot 2\pi$, marked by the gray dashed line). (b) Strength $J$ of the effective $\textit{XX}$ coupling. Device parameters can be found in Tab.~\ref{table:params}.}
\end{figure}

We compute the low-energy spectra of the full-model Hamiltonian $H$ as well as the effective Hamiltonian $H_{\mathrm{eff}}$ and plot the results in Fig.~\ref{fig:spectrum}(a). We vary the coupler flux along the contour $\mathcal{C}$ shown in Fig.~\ref{fig:flux_contour} to ensure that the qubits remain at their sweet spots. Relative deviations between the two spectra are at the level of a percent or less, indicating that the exact results can be accurately described by an effective model of two qubits coupled by a tunable $\textit{XX}$ interaction. The value of the tunable-coupling coefficient $J$ is shown in Fig.~\ref{fig:spectrum}(b) and crosses through zero at $\phi_{c}\approx0.27\cdot2\pi$. At this position in flux space, the coupler is in the ``off" state.

To quantify the on-off ratio of the tunable coupler, we numerically calculate the strength of the parasitic $\textit{ZZ}$ interaction
using the formula $\zeta_{ZZ}=E_{\overline{1100}}-E_{\overline{1000}}-E_{\overline{0100}}+E_{\overline{0000}}$ \cite{Ficheux2021}. The eigenenergy $E_{\overline{ijnp}}$ of the dressed state $|\overline{i_{a}, j_{b}, n_{-}, p_{+}}\rangle$ is found by numerically diagonalizing the full model Hamiltonian $H$. The $\textit{ZZ}$ interaction strength $\zeta_{ZZ}$ is less than $0.3$ $h\cdot$kHz at the off position for the parameters considered here, implying an on-off ratio on the order of $10^5$. It is a general feature of coupled systems of low-frequency fluxonia that $\textit{ZZ}$ interaction strengths are suppressed, due to the small repulsions between computational and non-computational states \cite{Ficheux2021}.

\section{Single-Qubit Gates}
\label{sec:single_q_gates}

There are important differences between how single-qubit gates are performed on high frequency qubits like transmons and how they are executed on low-frequency qubits like those studied here. For transmon qubits, drive strengths are typically small compared with the qubit frequency. It is then appropriate to move into a frame co-rotating with the drive frequency (typically on or near resonance with the qubit frequency) and perform the RWA \cite{Krantz2019, Blais2021}. The rotating-frame Hamiltonian is now time independent, allowing for the relatively straightforward calculation of time-evolution operators (propagators). Observe that in this rotating frame, idling corresponds to an identity operation (assuming a resonant drive). 
In contrast, to obtain fast gates for low-frequency qubits like heavy fluxonium \cite{Zhang2020} or superconducting composite qubits \cite{Campbell2020}, drive strengths typically equal or exceed the qubit frequencies. Thus, gates are typically performed in the laboratory frame as it is not appropriate to move into a rotating frame like that described above \cite{Zhang2020, Campbell2020, Yang2017, Wu2007}. 
In the lab frame, qubit states acquire dynamical phase factors while idling. Indeed we utilize these $\textit{Z}$ rotations in Sec.~\ref{sec:two_q_gate} for achieving a high-fidelity $\siswap$ gate. Nevertheless in the absence of drives, we obtain an identity operation (up to an overall sign) only by idling for exact multiples of the Larmor period $\tau_{q}=2\pi/\omega_{q}$, where $\omega_{q}$ is the qubit frequency. If we now consider multiple qubits with non-commensurate frequencies, it is not obvious how to perform an operation on one qubit without a second qubit acquiring dynamical phase during the gate time. 
Therefore, we seek an active means of obtaining {\it variable-time} identity operations for low-frequency qubits.
Single-qubit $\textit{X}/2$ and $\textit{Y}/2$ gates can be obtained using the techniques described in e.g. Refs.~\cite{Zhang2020, Campbell2020}, allowing for universal control when combined with arbitrary $\textit{Z}$ rotations achieved by idling.

We utilize flux pulses that begin and end at zero and whose shapes are described by sinusoidal functions, but that only last for a single period \cite{Campbell2020}. This pulse shape is chosen because the external flux averages to zero \footnote{Many other simple pulse shapes achieve net-zero flux, such as those utilized in Ref.~\cite{Zhang2020},
and can yield high-fidelity gates. Single-period sinusoids are used here for simplicity.
}, helping eliminate long-timescale distortions \cite{Rol2019}. The Hamiltonian of a single fluxonium biased at the half-flux sweet spot and subject to a sinusoidal flux drive $\phi(t)=\delta\phi\sin(\omega_{d}t)$ is $H_{\mathrm{fl}}(t)=H_{\pi}+H_{d}(t)$, where
\begin{align}
\label{eq:fluxonium}
H_{\pi} &=4E_{C}n^{2}-E_{J}\cos(\varphi)+\frac{1}{2}E_{L}(\varphi+\pi)^{2}, \\ 
H_{d}(t) &= E_{L}\varphi\delta\phi\sin(\omega_{d}t).
\end{align}
Projecting onto the computational subspace yields \cite{Zhang2020}
\begin{align}
\label{eq:fluxoniumspin}
H_{q}(t) = -\frac{\omega_{q}}{2}\sigma_{z}+A\sin(\omega_{d}t)\sigma_{x},
\end{align}
defining the effective drive amplitude $A=E_{L}\langle 0 | \varphi | 1 \rangle\delta\phi$ and making use of selection rules at the half-flux sweet spot.
For typical heavy-fluxonium parameters such as those chosen for qubits $a$ and $b$, the amplitude of the drive $A$ exceeds the qubit frequency $\omega_{q}$ for deviations from the sweet spot as small as $\delta\phi=0.02\cdot 2\pi$. Indeed, such strong drives have been used to implement fast single-qubit gates with high fidelities \cite{Zhang2020,Campbell2020}. Here, we utilize similarly strong drives for the implementation of identity pulses. We seek conditions on the drive strength $A$ and frequency $\omega_{d}$ such that the propagator $U_{q}(t)$ is equal to the identity operation after a single drive period, $U_{q}(t=2\pi/\omega_{d})=\openone$. The propagator satisfies the time-dependent Schr\"odinger equation
\begin{align}
\frac{d U_{q}(t)}{dt}=-i H_{q}U_{q}(t),
\end{align}
with the initial condition $U_{q}(0)=\openone$. In the regime where the qubit frequency $\omega_{q}$ is small compared to the drive amplitude $A$, it is appropriate to move into the interaction picture defined with respect to the drive. This transformation is achieved via the unitary
\begin{align}
U_{0}(t)&=\exp\left(-i\int_{0}^{t}dt'\,A\sin[\omega_{d}t]\sigma_{x}\right) \\ \nonumber 
&=\exp\left(-i\frac{2A}{\omega_{d}}\sin^2\left[\frac{\omega_{d}t}{2}\right]\sigma_{x}\right).
\end{align}
The interaction-frame Hamiltonian is
\begin{align}
H_{q}'&=U_{0}^{\dagger}H_{q}U_{0}-iU_{0}^{\dagger}\dot{U}_{0} \\ \nonumber
&=-\frac{\omega_{q}}{2}\cos\left(\frac{4A}{\omega_{d}}\sin^{2}\left[\frac{\omega_{d}t}{2}\right]\right)\sigma_{z} \\ \nonumber 
&\quad-\frac{\omega_{q}}{2}\sin\left(\frac{4A}{\omega_{d}}\sin^{2}\left[\frac{\omega_{d}t}{2}\right]\right)\sigma_{y},
\end{align}
while the propagator in this frame $U_{q}'(t)$ satisfies $d U_{q}'(t)/dt=-i H_{q}'U_{q}'(t)$. To obtain an approximation to the propagator $U_{q}'(t)$ we carry out a Magnus expansion \cite{Magnus, Blanes2009, Wilcox} in which the propagator is assumed to take an exponential form $U_{q}'(t)=\exp(\sum_{i=1}^{\infty}\Delta_{i}(t))$. The first term in this series is $\Delta_{1}(t)=-i\int_{0}^{t}H'_{q}(t')dt'$, and the formulas for terms up to fourth order are given in e.g. Refs.~\cite{Blanes2009, Wilcox}. We truncate the Magnus series after the first term, as we find that higher-order terms are generally small and can be neglected. The propagator at the conclusion of the pulse is
\begin{align}
\nonumber 
U_{q}'(\tau_{d})
&=\cos\left(\frac{\pi \omega_{q}}{\omega_{d}}J_{0}\left[\frac{2A}{\omega_{d}}\right]\right)\openone
+ i\sin\left(\frac{\pi \omega_{q}}{\omega_{d}}J_{0}\left[\frac{2A}{\omega_{d}}\right]\right)
\\ 
\label{eq:Uq}
&\quad\times \left(\cos\left[\frac{2A}{\omega_{d}}\right]\sigma_{z}+\sin\left[\frac{2A}{\omega_{d}}\right]\sigma_{y}\right),
\end{align} 
where $\tau_{d}=2\pi/\omega_{d}$ and $J_{0}$ is the zeroth-order Bessel function of the first kind. The propagators in the lab and interaction frames are related by $U_{q}(t)=U_{0}(t)U_{q}'(t)U_{0}^{\dagger}(0)$. Because the lab and interaction frames coincide at $t=0$ and $t=\tau_{d}$, the propagators in the lab and interaction frames are the same at the conclusion of the pulse. To obtain an identity gate, the general solution is
\begin{align}
\label{eq:gensol}
\frac{\pi \omega_{q}}{\omega_{d}}J_{0}\left(\frac{2A}{\omega_{d}}\right)=2\pi r, \quad r\in \mathds{Z},
\end{align}
which is an equation for the variables $A, \omega_{d}$. Solutions which avoid fixing $\omega_{d}$ based on the value of $\omega_{q}$ are those for $r=0$ which satisfy
\begin{align}
\label{eq:Bessel}
\frac{2A}{\omega_{d}}=j_{k}, \quad k=1,2,\ldots,
\end{align}
where $j_{k}$ is the $k^{\mathrm{th}}$ zero of $J_{0}$. Thus, by choosing a combination of drive amplitude $A$ and frequency $\omega_{d}$ (and thus gate time) obeying Eq.~\eqref{eq:Bessel}, we obtain a variable-time identity gate. We note that it is also possible to arrive at Eq.~\eqref{eq:gensol} using a perturbative analysis in the context of Floquet theory \cite{Huang2021}.

\begin{figure}
    \centering
    \includegraphics[width=\columnwidth]{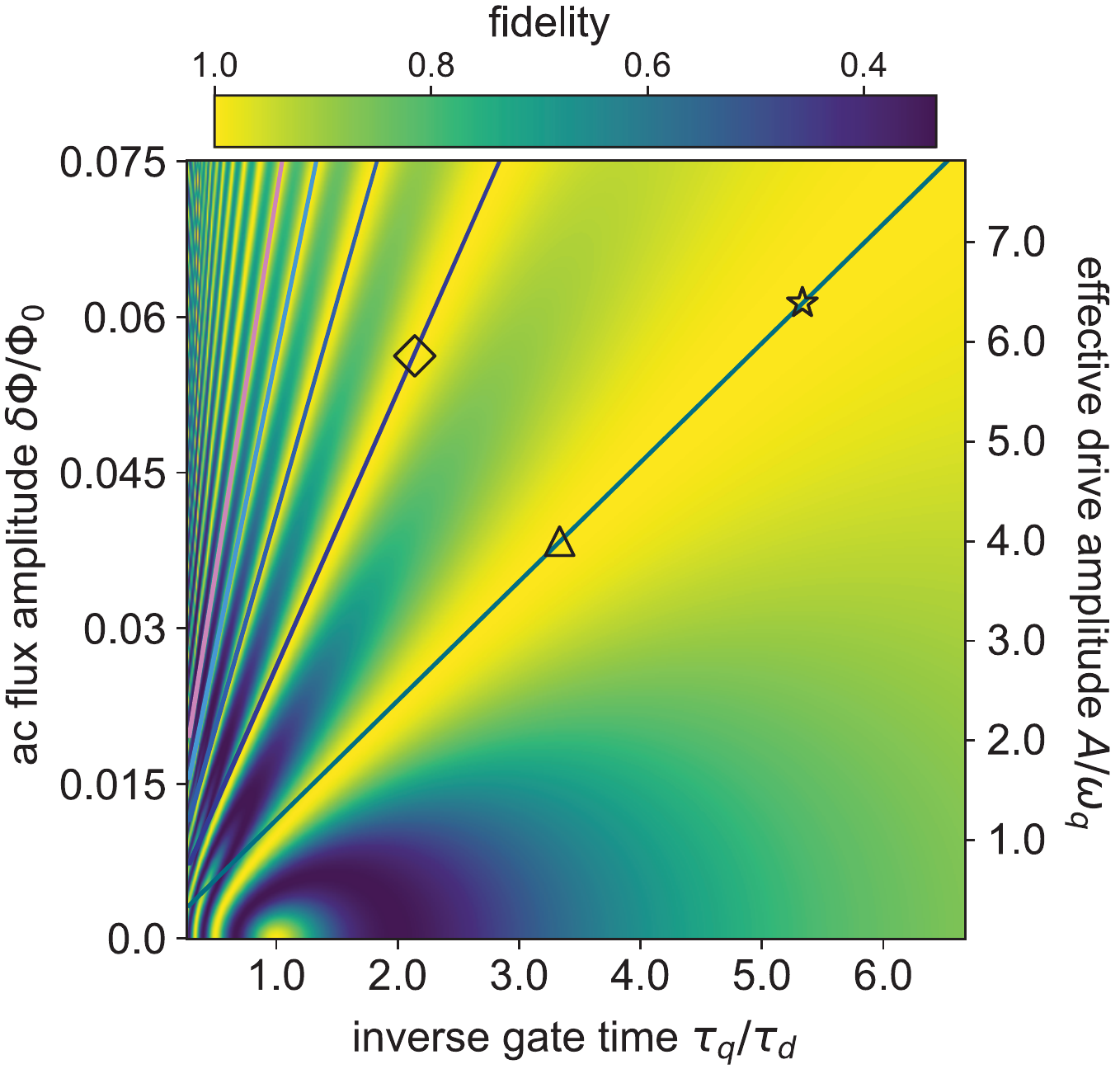}
    \caption{\label{fig:id} Variable-time single-qubit identity pulse. We plot the fidelity of an identity operation under a single-period sinusoidal qubit-flux drive. Lines mark locations in amplitude and frequency space where $A=j_{k}\omega_{d}/2$ for the first 5 zeroes $j_{k}$ of $J_{0}$. Multiple other lines of high fidelity corresponding to larger zeros of $J_{0}$ are visible in the numerics. These variable-time identity gates are ultra-fast, with gate times that can be small compared to the Larmor period.
    The point-like region of high fidelity at $\delta\phi=0$ corresponds to the passive identity operation. The marked points label example drive parameters used for visualizing Bloch sphere trajectories in Fig.~\ref{fig:idBloch}. Numerical simulations are performed using the parameters of qubit $b$ where $\omega_{q}/2\pi=37$ MHz.}
\end{figure}

We present numerical results illustrating that the proposed identity gates can be achieved with high fidelity. To calculate the closed-system fidelity of a quantum operation we utilize the formula
\cite{Pedersen2007}
\begin{align}
\label{eq:F}
F=\frac{\Tr(U^{\dagger}U)+|\Tr[U_{\mathrm{T}}^{\dagger}U]|^2}{d(d+1)},
\end{align}
where $d$ is the dimension of the relevant subspace of the Hilbert space, $U_{\mathrm{T}}$ is the target unitary and $U$ is the projection onto the $d$ dimensional subspace of the propagator realized by time evolution. This formula is especially useful when considering systems where leakage may be an issue; in such cases, deviations of the operator $U$ from unitarity are penalized by the term $\Tr(U^{\dagger}U)$. To obtain the propagator associated with time evolution under the Hamiltonian $H_{\mathrm{fl}}(t)$, it is most appropriate to express $H_{\mathrm{fl}}(t)$ in the eigenbasis of the static Hamiltonian $H_{\pi}$. The qubit states are the two lowest-energy states, and we retain up to eight eigenstates to monitor leakage. Diagonalization of $H_{\pi}$ is done using scqubits \cite{scqubits}, while time-dependent simulations are performed using QuTiP \cite{qutip1, *qutip2}. 
Sweeping over the drive frequency and amplitude of the flux pulse, we monitor the fidelity of an identity operation, taking $d=2$ and $U_{T}=\openone$ in Eq.~\eqref{eq:F}, see Fig.~\ref{fig:id}.
\begin{figure}
    \centering
    \includegraphics[width=0.9\columnwidth]{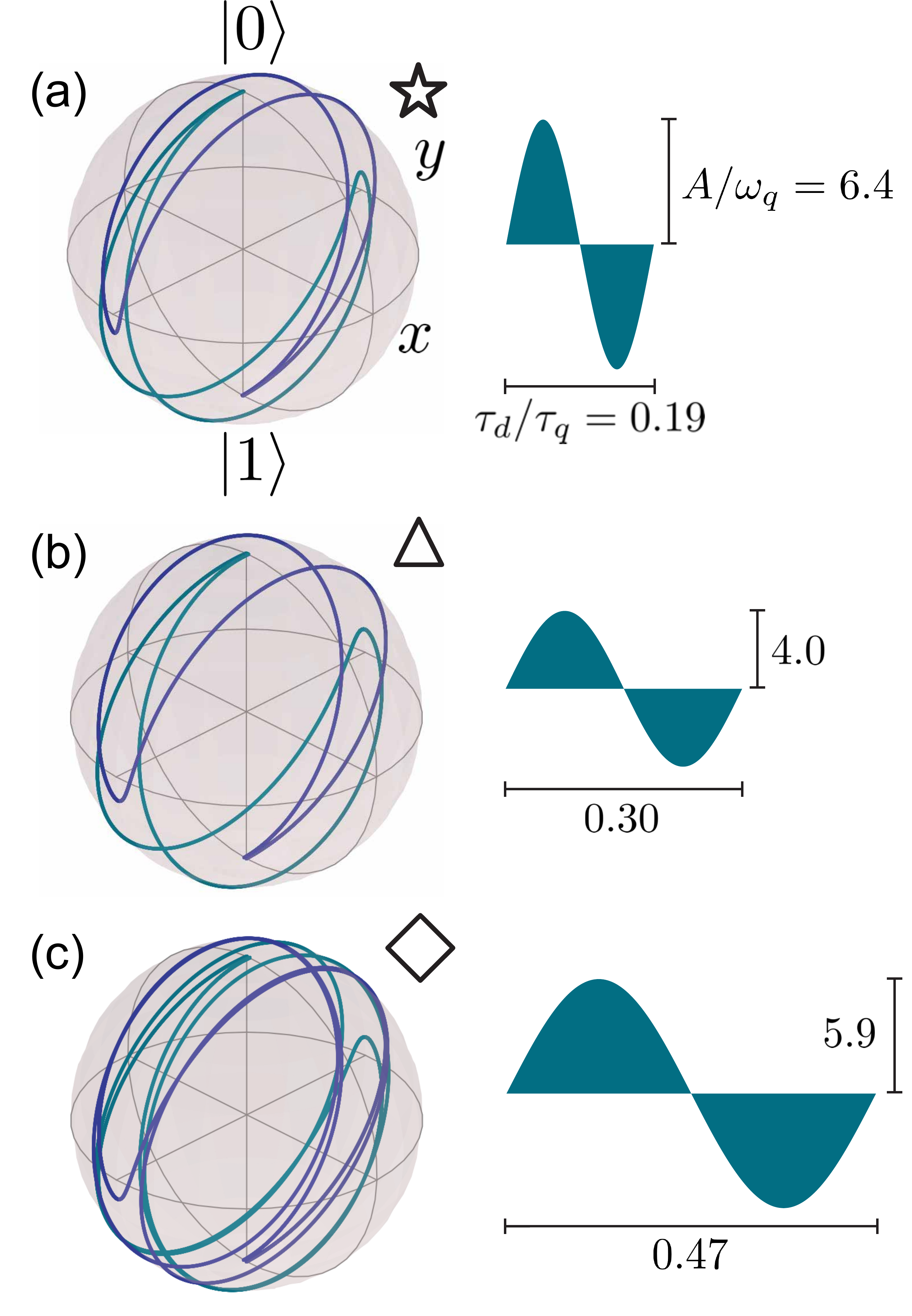}
    \caption{\label{fig:idBloch} (a-c) Bloch sphere trajectory in the lab frame of the initial states $|0\rangle$ and $|1\rangle$ subject to the pictured identity gates. The pulse parameters utilized here are marked in Fig.~\ref{fig:id}. Each gate achieves an identity operation with fidelity $F\geq0.9998$. }
\end{figure}
Regions of high fidelity appear as ``fingers" in the space of inverse gate time (drive frequency $\omega_{d}$) and effective drive amplitude $A$. The colored lines are given by $A=j_{k}\omega_{d}/2$ for $k=1,2,3,4,5$, corresponding to the drive parameters that analytically predict identity gates. These lines overlap with the regions of high fidelity computed numerically for large amplitude $A$ compared with the qubit frequency $\omega_{q}$. For decreasing $\omega_{d}$ and $A$, the lines begin to deviate from the high-fidelity fingers due to the breakdown of the Magnus expansion \cite{Blanes2009}. Nevertheless, we find numerically that high-fidelity $F>0.9999$ identity gates can be achieved across a wide range of inverse gate times 
$0.5\lesssim\omega_{d}/\omega_{q}\lesssim7$.
Leakage outside the computational subspace is negligible for the parameters considered here.

The time evolution of the qubit states in the lab frame in the form of trajectories on the Bloch sphere during identity pulses is shown in Fig.~\ref{fig:idBloch}. The drive frequencies $\omega_{d}$ used in Figs.~\ref{fig:idBloch}(a)-(c) are 
$\omega_{d}/\omega_{q}=5.3, 3.3, 2.1$, with drive amplitudes $A$ obtained from Eq.~\eqref{eq:Bessel} using the Bessel zeros $j_{1}, j_{1}, j_{2}$,  respectively.
In all three cases we obtain fidelities of $F\geq0.9998$. 
Numerical optimization of the drive amplitude keeping the drive frequency fixed yields $F>0.99999$ in each case. The optimized amplitude generally differs from the amplitude derived analytically by less than or of the order of a percent.

In this section, we analyzed an isolated heavy fluxonium subject to an ac flux drive. 
The generalization to fluxonia embedded in the architecture considered in this work is straightforward. At the off position, the operator activated by a flux drive on qubit $a$ and $b$ is to a good approximation of the form of $\textit{XI}$ and $\textit{IX}$, respectively, see Appendix~\ref{appendix:tdH} for details. Thus, the identity-gate protocol is immediately applicable to each individual qubit. 

\section{Two-qubit entangling gate}
\label{sec:two_q_gate}

When performing two-qubit gates on low-frequency fluxonium qubits, we encounter similar issues to those present for single-qubit gates. To achieve relatively fast gate times compared with $T_{1}$ and $T_{2}$, we utilize drive strengths where the RWA is invalid. We perform a Magnus expansion in a frame co-rotating with the qubit frequencies to account for the counter-rotating terms order-by-order. We note that similar results can be obtained using a Floquet analysis \cite{Shirley1965, Petrescu2021}. Because single-qubit gates are performed in the lab frame, we then transform the rotating-frame propagator back into the lab frame.

To activate a two-qubit interaction, we consider an ac sinusoidal coupler flux drive $\phi_{c}(t)=\delta\phi_{c}\sin(\omega_{d}t)$. 
The time-dependent Hamiltonian in the computational subspace is 
\begin{align}
\label{eq:Ht2qmaintext}
H_{2q}(t)=-\frac{\omega_{a}'}{2}\bar{\sigma}_{z}^{a}-\frac{\omega_{b}'}{2}\bar{\sigma}_{z}^{b}+A\sin(\omega_{d}t)\bar{\sigma}_{x}^{a}\bar{\sigma}_{x}^{b},
\end{align}
where we emphasize that gates are performed in the basis of the dressed states, 
see Appendix~\ref{appendix:tdH} for details. The Pauli matrices are defined as e.g. $\bar{\sigma}_{x}^{a}=\sum_{j=0,1}\oket{0j}\obra{1j}+\mathrm{H.c.}$, etc., utilizing the shorthand 
$\oket{i_{a}j_{b}}\equiv\oket{i_{a},j_{b},0_{-},0_{+}}$. The effective drive amplitude $A\equiv J_{\mathrm{ac}}\delta\phi_{c}$ is defined in terms of the ac two-qubit coupling strength $J_{\mathrm{ac}}$ given in Eq.~\eqref{eq:Jac}. For our parameters, $J_{\mathrm{ac}}=18.3$ MHz. 
Just as for single-qubit gates, we drive with $\sin\omega_{d}t$ rather than $\cos\omega_{d}t$ because we intend to activate the interaction for only one or a few drive periods $n$ \footnote{The reason we keep the treatment general enough to include multiple drive periods will become clear when we consider time evolution on the full system: for realistic parameters, a single drive period leads to drive amplitudes so large that we obtain fidelity-degrading contributions from high-lying coupler states.}. 
In general, the propagator at the final time $\tau_{n}=2\pi n/\omega_{d}$ is
\begin{align}
U(\tau_{n})=\mathcal{T}e^{-i\int_{0}^{\tau_{n}}H_{2q}(t')dt'}=\left(\begin{matrix}
a & 0 & 0 & d \\
0 & b & c & 0 \\
0 & -c^{*} & b^{*} & 0 \\
-d^{*} & 0 & 0 & a^{*}
\end{matrix} \right),
\end{align}
where $\mathcal{T}$ is the time-ordering operator and $|a|^{2}+|d|^{2}=|b|^{2}+|c|^{2}=1$. To obtain an entangling gate, we target drive parameters $A,\omega_{d}$ that yield $|b|=|c|=1/\sqrt{2}$ and $d=0$ \footnote{A gate with $|a|=|d|=1/\sqrt{2}$ and $|c|=0$ is also entangling, yielding a $\sbswap$-like gate \cite{Poletto2012}. Here, we focus instead on entangling gates performed in the $\oket{01}, \oket{10}$ subspace rather than in the $\oket{00}, \oket{11}$ subspace.}. We parametrize this gate as 
\begin{align}
\label{eq:sphiswap}
\sphiswap=\left(\begin{matrix}
e^{i\alpha} & 0 & 0 & 0 \\
0 & e^{i\beta}/\sqrt{2} & e^{i\gamma}/\sqrt{2} & 0 \\
0 & -e^{-i\gamma}/\sqrt{2} & e^{-i\beta}/\sqrt{2} & 0 \\
0 & 0 & 0 & e^{-i\alpha}
\end{matrix}\right).
\end{align}
To see that this gate is entangling, note either that it can produce Bell states or that it can be transformed into the entangling gate
\begin{align}
\siswap = \left(\begin{matrix}
1 & 0 & 0 & 0 \\
0 & 1/\sqrt{2} & -i/\sqrt{2} & 0 \\
0 & -i/\sqrt{2} & 1/\sqrt{2} & 0 \\
0 & 0 & 0 & 1
\end{matrix}\right),
\end{align}
using only single-qubit operations \cite{Vedral1997}. One such transformation using single-qubit gates is
\begin{align}
\label{eq:RZ_phiswap}
\siswap=R_{Z}^{a}(\theta_{a1})R_{Z}^{b}(\theta_{b1})\sqrt{\phi\mathrm{SWAP}}R_{Z}^{a}(\theta_{a2})R_{z}^{b}(\theta_{b2}),
\end{align}
where 
\begin{align*}
R_{Z}^{j}(\theta)=\exp(-i\theta\bar{\sigma}_{z}^{j}/2),\quad j=a,b.
\end{align*}
Expressions for the $\textit{Z}$ rotation angles in Eq.~\eqref{eq:RZ_phiswap} in terms of $\alpha,\beta$ and $\gamma$ are specified below. The relationship \eqref{eq:RZ_phiswap} provides an explicit recipe for constructing a $\siswap$ gate, given a $\sphiswap$ gate and arbitrary single-qubit $\textit{Z}$ rotations. Quantum algorithms are typically written in terms of named gates like $\siswap$ \cite{siswap1, *siswap2, *siswap3}, as opposed to the native gate $\sphiswap$ achieved here. Thus, it may be useful to immediately transform the obtained $\sphiswap$ gate into the more familiar $\siswap$. This is the strategy we pursue here. 

Generally, only three of the $\textit{Z}$ rotations in Eq.~\eqref{eq:RZ_phiswap} are necessary. We make use of the freedom of the extra $\textit{Z}$ rotation by choosing the angle $\theta_{b2}\in[0,2\pi)$ that optimizes the overall gate time, including the $\textit{Z}$ rotations. The remaining angles are set to
\begin{align}
\label{eq:Zrotangles}
\theta_{a1} &= \frac{\pi}{2} + \alpha + \gamma -\theta_{b2}, \\ \nonumber 
\theta_{b1} &= \alpha-\beta-\theta_{b2}, \\ \nonumber 
\theta_{a2} &= -\frac{\pi}{2} +\beta -\gamma +\theta_{b2},
\end{align}
to satisfy Eq.~\eqref{eq:RZ_phiswap}.
In the following, we find explicit expressions for $\alpha, \beta$ and $\gamma$ in terms of the drive parameters and qubit frequencies. Because we operate in the lab frame, these $\textit{Z}$ rotations are obtained by idling. Idle times for coincident $\textit{Z}$ rotations may differ in general, therefore to synchronize the time spent performing single-qubit gates we make use of the variable-time single-qubit identity gates discussed in Sec.~\ref{sec:single_q_gates}.

\subsection{Constructing \texorpdfstring{$\sphiswap$}{}}

The propagator $\sphiswap$ can be obtained from time evolution under the Hamiltonian $H_{2q}(t)$ as follows. The qubit frequencies $\omega_{a}',\omega_{b}'$ are fixed by operating the qubits at their sweet spots, while the drive parameters $A, \omega_{d}$ may be varied. 
The Hamiltonian $H_{2q}(t)$ only couples the pairs of states $|\overline{00}\rangle\leftrightarrow|\overline{11}\rangle$, $|\overline{01}\rangle\leftrightarrow|\overline{10}\rangle$, thus $H_{2q}(t)$ decomposes into a direct sum $H_{2q}(t)=H_{-}(t)\bigoplus H_{+}(t)$, where
\begin{align}
H_{\pm}(t)=-\frac{\omega_{\pm}}{2}\Sigma_{z}^{\pm}+A\sin(\omega_{d}t)\Sigma_{x}^{\pm},
\end{align}
defining $\omega_{\pm}=\omega_{a}'\pm\omega_{b}'$.
The Hamiltonians $H_{+}(t), H_{-}(t)$ describe dynamics in the $\oket{00}, \oket{11}$, and the $\oket{01},\oket{10}$ subspaces, respectively. The corresponding Pauli matrices are denoted by $\Sigma_{j}^{\pm}$, for example $\Sigma_{z}^{+}=|\overline{00}\rangle\langle\overline{00}|-|\overline{11}\rangle\langle\overline{11}|$.
For realistic parameters, the two-level-system frequencies $\omega_{+}$ and $\omega_{-}$ are large compared with the drive amplitude $A$. In this case it is appropriate to move into the interaction picture defined by the unitaries
\begin{align}
U_{0}^{\pm}(t)=\exp\left[i\frac{\omega_{\pm}}{2}\Sigma_{z}^{\pm}t \right].
\end{align}
The interaction-frame Hamiltonians are
\begin{align}
\label{eq:Hpm}
H_{\pm}'(t) = A\sin(\omega_{d}t)[\cos(\omega_{\pm} t)\Sigma_{x}^{\pm}+\sin(\omega_{\pm} t)\Sigma_{y}^{\pm}].
\end{align}

To calculate the associated propagators, we carry out a Magnus expansion including the first- and second-order terms. It is straightforward to calculate higher-order corrections, however we find for our parameters that they are small and can be neglected. The expression for the propagator is then $U_{\pm}'(t)=\exp(\Delta_{1}^{\pm}[t]+\Delta_{2}^{\pm}[t])$, where \cite{Wilcox, Blanes2009, Magnus}
\begin{align}
\Delta_{1}^{\pm}(t) &=-i\int_{0}^{t}H_{\pm}'(t')dt', \\ 
\Delta_{2}^{\pm}(t) &=-\frac{1}{2}\int_{0}^{t}dt_{1}\int_{0}^{t_{1}}dt_{2}[H_{\pm}'(t_{1}),H_{\pm}'(t_{2})].
\end{align}
At the conclusion of the gate $t=\tau_{n}$, we obtain
\begin{align}
\label{eq:propsmall}
U_{\pm}'(\tau_{n}) &=\cos\left(\xi_{\pm}\right)\openone -i\sin\left(\xi_{\pm}\right)(\vec{n}_{\pm}\cdot\vec{\Sigma}^{\pm}),
\end{align}
where 
\begin{align}
\vec{n}_{\pm}&=(\sin[\pi n \omega_{\pm}/\omega_{d}],-\cos[\pi n \omega_{\pm}/\omega_{d}],\varepsilon_{\pm}/\xi_{\pm}), \\ 
\vec{\Sigma}^{\pm}&=\left( \Sigma_{x}^{\pm}, \Sigma_{y}^{\pm}, \Sigma_{z}^{\pm} \right),
\end{align}
and we have defined
\begin{align}
\xi_{\pm} &=\frac{2A\omega_{d}}{\omega_{d}^2-\omega_{\pm}^2}\sin\left(\frac{\pi n \omega_{\pm}}{\omega_{d}}\right),\\ \nonumber 
\varepsilon_{\pm} &=\frac{A^2\omega_{d}^2\sin(2\pi n \omega_{\pm}/\omega_{d})}{(\omega_{d}^2-\omega_{\pm}^2)^2}+\frac{A^2\pi n \omega_{\pm}}{\omega_{d}(\omega_{d}^2-\omega_{\pm}^2)}.
\end{align}
We have neglected corrections of order $\mathcal{O}([A/\omega_{d}]^3)$ to $\xi_{\pm}$. 
The first-order terms involving $\Sigma_{x}^{\pm}, \Sigma_{y}^{\pm}$ determine the amount of population transfer between the two states in each subspace. The second-order terms encode the leading-order beyond-the-RWA corrections and are proportional to $\Sigma_{z}^{\pm}$. Indeed, in the resonant limit, the first-order terms $\Delta_{1}^{\pm}(\tau_{n})=-i\tau_{n}\frac{A}{2}\Sigma_{y}^{\pm}$ reproduce the RWA results \cite{Blais2021} while the second-order terms $\Delta_{2}^{\pm}(\tau_{n})=i\tau_{n}\frac{3A^{2}}{8\omega_{d}}\Sigma_{z}^{\pm}$ correspond to the well-known Bloch-Siegert shift \cite{Zeuch2020, BlochSiegert}.
Transforming the propagator back to the lab frame via the identity $U_{\pm}(\tau_{n})=U_{0}^{\pm}(\tau_{n})U_{\pm}'(\tau_{n})U_{0}^{\pm}(0)^{\dagger}$, we find
\begin{align}
\nonumber
U_{\pm}(\tau_{n})&=\exp(i\vartheta_{\pm}\Sigma_{z}^{\pm})[\cos(\xi_{\pm})\openone-i\varepsilon_{\pm}\sinc(\xi_{\pm})\Sigma_{z}^{\pm}] \\
\label{eq:labframeprops}
&+i\sin(\xi_{\pm})\Sigma_{y}^{\pm}\\ \nonumber 
\approx &\cos(\xi_{\pm})\exp\{i[\vartheta_{\pm}-\varepsilon_{\pm}\tanc(\xi_{\pm})]\Sigma_{z}^{\pm}\}\\ \nonumber &+i\sin(\xi_{\pm})\Sigma_{y}^{\pm},
\end{align}
defining $\vartheta_{\pm}=\pi n \omega_{\pm}/\omega_{d}$. The approximate equality is valid for $\tanc(\xi_{\pm})\varepsilon_{\pm}\ll1$, and $\tanc(x)=\tan(x)/x$. 

\subsection{Determining optimal drive parameters}

To obtain the $\sphiswap$ gate, we require
\begin{widetext}
\begin{align}
\label{eq:Up}
U_{+}(\tau_{n})&=\left(\begin{matrix}
\cos(\xi_{+})e^{i(\vartheta_{+}-\tanc[\xi_{+}]\varepsilon_{+})} & \sin(\xi_{+}) \\ -\sin(\xi_{+}) & \cos(\xi_{+})e^{-i(\vartheta_{+}-\tanc[\xi_{+}]\varepsilon_{+})}
\end{matrix}
\right) 
\stackrel{!}{=} \left(\begin{matrix}
e^{i\alpha} & 0 \\ 
0 & e^{-i\alpha}
\end{matrix} \right), \\
\label{eq:Um}
U_{-}(\tau_{n}) &=\left(\begin{matrix}
\cos(\xi_{-})e^{i(\vartheta_{-}-\tanc[\xi_{-}]\varepsilon_{-})} & \sin(\xi_{-}) \\ -\sin(\xi_{-}) & \cos(\xi_{-})e^{-i(\vartheta_{-}-\tanc[\xi_{-}]\varepsilon_{-})}
\end{matrix}
\right) 
\stackrel{!}{=}
\frac{1}{\sqrt{2}}
\left(\begin{matrix}
e^{i\beta} & e^{i\gamma} \\ 
-e^{-i\gamma} & e^{-i\beta}
\end{matrix} \right).
\end{align}
\end{widetext}

The solution for Eq.~\eqref{eq:Up} is
\begin{align}
\label{eq:plus_cond}
\xi_{+}=\frac{2A\omega_{d}}{\omega_{d}^2-\omega_{+}^2}\sin\left(n\pi\frac{\omega_{+}}{\omega_{d}}\right)=p\pi,\quad p\in\mathds{Z},
\end{align}
which should be interpreted as an equation involving the unknowns $A, \omega_{d}$. For any nonzero $A$, solutions to Eq.~\eqref{eq:plus_cond} for $p=0$ are
\begin{align}
\label{eq:omegad}
\omega_{d}=n\omega_{+}/m,\quad (m=1,2,3,\ldots,m\neq n).
\end{align}
For nonzero $p$, solutions $(A,\omega_{d})$ can only be found by numerically solving the full transcendental equation \eqref{eq:plus_cond}. We find in the following that to satisfy Eq.~\eqref{eq:Um}, it is necessary to have the freedom of varying the drive amplitude $A$. Thus, we only consider the case $p=0$. Setting $m=n$ is excluded in Eq.~\eqref{eq:omegad} as in this case the left-hand side of Eq.~\eqref{eq:plus_cond} does not vanish. However, this restriction is no issue, as motivated by the drive frequency $\omega_{d}=\omega_{-}$ used to obtain the $\siswap$ gate when the RWA is valid \cite{Blais2021} we do not consider on resonance driving of the $\oket{00}\leftrightarrow\oket{11}$ transition $\omega_{d}=\omega_{+}$. 
With $\omega_{d}$ given by Eq.~\eqref{eq:omegad}, the expression for $\varepsilon_{+}$ simplifies to $\varepsilon_{+}=\frac{A^2}{\omega_{d}^2}\frac{\pi n m}{1-m^2}$ and we satisfy Eq.~\eqref{eq:Up} with the phase $\alpha=\vartheta_{+}-\varepsilon_{+}$. 

Considering now the requirement \eqref{eq:Um} for $U_{-}$, the solution is
\begin{align}
\label{eq:xim}
\xi_{-}=\frac{2A\omega_{d}}{\omega_{d}^2-\omega_{-}^2}\sin\left(n\pi\frac{\omega_{-}}{\omega_{d}} \right)=\pm\frac{\pi}{4}+\pi q,\quad q\in \mathds{Z},
\end{align}
where the $\pm$ indicates that the sign may be absorbed into the phases $\beta, \gamma$. We interpret Eq.~\eqref{eq:xim} as an equation for the unknown $A$, as we have fixed $\omega_{d}$ previously. Solving for $A$ yields
\begin{align}
\label{eq:An}
A=\pm\frac{\pi(\omega_{d}^2-\omega_{-}^2)}{8\omega_{d}\sin(\pi n\frac{\omega_{-}}{\omega_{d}})},
\end{align}
where we have set $q=0$ to minimize the magnitude of $A$. In general, the fraction on the right-hand side of Eq.~\eqref{eq:An} may be positive or negative, depending on $n$ and the magnitude of $\omega_{-}$ relative to $\omega_{d}$. Thus, we choose the sign of $\xi_{-}=\pm\pi/4$ based on which yields a positive drive amplitude $A$. With the drive frequency and amplitude given by Eq.~\eqref{eq:omegad} and Eq.~\eqref{eq:An} respectively, we satisfy Eq.~\eqref{eq:Um} 
with phases $\beta=\vartheta_{-}-4\varepsilon_{-}/\pi$ and $\gamma=0$ or $\gamma=\pi$ depending on the sign of $\xi_{-}$. 

The previous analysis only leaves us to choose the integers $m, n$, see Eq.~\eqref{eq:omegad}. We make use of this freedom to limit the drive amplitude $A$ in magnitude. Careful inspection of the removable singularity in Eq.~\eqref{eq:An} suggests the usage of a drive frequency $\omega_{d}$ near $\omega_{-}$. This can be achieved by a combination of $n$ and $m$ such that their ratio $n/m$ closely approximates $\omega_{-}/\omega_{+}$. The optimal choice of $n$ must balance between mitigating the effects of $T_{1}$ and $T_{2}$ by keeping gate durations $2\pi n/\omega_{d}$ as short as possible, and holding at bay unwanted population transfer incurred by strong drive amplitudes $A\sim1/n$, see Appendix~\ref{appendix:tdH} for details. With $n$ and $m$ specified as such, we have constructed the $\sphiswap$ gate allowing for the execution of a $\siswap$ gate when combined with single-qubit $\textit{Z}$ rotations. 

\subsection{Full-system numerical simulations}

This realization of $\siswap$ reaches closed-system (open-system) fidelities as high as $F=0.9996$ ($F=0.9994$), which we obtain from numerical simulation of the full system as detailed in the following. Time evolution is based on the Hamiltonian $H(t)=H_{0}+V+\sum_{\mu=a,b,c}h_{\mu}\delta\phi_{\mu}(t)$, see Eqs.~\eqref{eq:H0}-\eqref{eq:V} as well as Eqs.~\eqref{eq:Ha}-\eqref{eq:Hc} for the flux-activated terms. The dc fluxes entering $H_{0}+V$ are set to the off position \footnote{The off position is found by minimizing the energy of the $|\overline{11}\rangle$ state, analogously to how the off position is found based on the effective Hamiltonian $H_{\mathrm{eff}}$. The exact and effective values for the coupler flux at the off position typically have a relative deviation of less than one percent}, as is appropriate for performing single- and two-qubit gates. For numerical efficiency, $H(t)$ is expressed in the eigenbasis of the static Hamiltonian $H_{0}+V$. The computational states of interest are the four lowest-energy states, with qubit frequencies $\omega_{a}'/2\pi=58.1$ MHz$,\omega_{b}'/2\pi=35.5$ MHz. Beyond these, we include up to 50 additional states in our simulations. For the parameters considered here, we have $\omega_{-}/\omega_{+}=0.24$, thus we choose $n/m=1/4$, yielding $\omega_{d}/2\pi=23.4$ MHz. Upon including the effects of decoherence, the choice $n=2$ and thus $m=8$, $A/2\pi=2.9$ MHz optimizes the gate fidelity.
\begin{figure}
    \centering
    \includegraphics[width=\columnwidth]{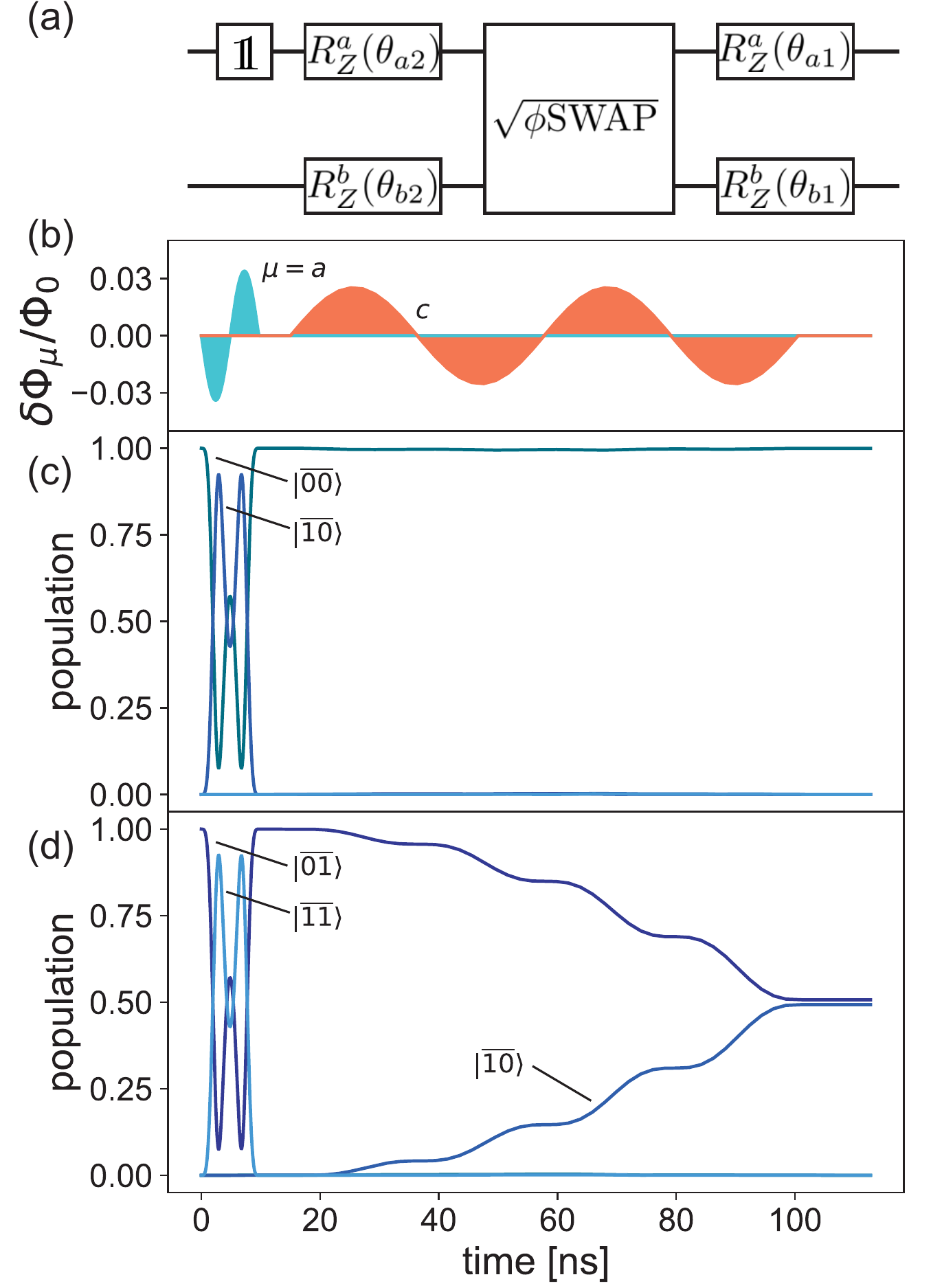}    \caption{\label{fig:sqrtiswap} $\siswap$ gate composed of a $\sphiswap$ gate and corrective $\textit{Z}$ rotations. (a) Quantum-circuit representation of the flux pulses shown in (b). The qubit flux $\delta\phi_{a}(t)$ (blue) is modulated to achieve the required identity operation, while the coupler flux $\delta\phi_{c}(t)$ (orange) is activated to entangle the qubits. Single-qubit $\textit{Z}$ rotations are obtained during the time spent idling. (c) Time evolution with the initial state $|\overline{00}\rangle$. Population transfer to the $|\overline{11}\rangle$ or other states is negligible during the $\sphiswap$ time window. (b) Time evolution for $|\overline{01}\rangle$ as the initial state. The $\oket{01}$ and $\oket{10}$ states exchange population during the segment when the coupler flux is nonzero. The closed-system fidelity of the $\siswap$ gate is $F=0.9996$.}
\end{figure}

The full gate duration is $t_{\mathrm{tot}}=2\pi n/\omega_{d}+\max(t_{a1},t_{b1})+\max(t_{a2},t_{b2})$, where $t_{\mu i}=-\theta_{\mu i}/\omega_{\mu}$. The equations for the times $t_{\mu i}$ are understood modulo $2\pi$ and the $\textit{Z}$ rotation angles are known in terms of the phases $\alpha,\beta,\gamma$, see Eq.~\eqref{eq:Zrotangles}.
The angle $\theta_{b2}$ is a free parameter and is chosen to minimize the overall gate time by forcing the idle times after the $\sphiswap$ gate to coincide $t_{a1}=t_{b1}$. For our parameters we obtain $t_{\mathrm{tot}}=113$ ns, where $2\pi n/\omega{d}=85$ ns and the single-qubit gates require 28 ns, see Fig.~\ref{fig:sqrtiswap}. For the initial state $\oket{00}$, population appreciably varies only during the single-qubit identity-gate segments, see Fig.~\ref{fig:sqrtiswap}(c). Meanwhile, for the state $\oket{01}$, population transfer to the $\oket{10}$ state occurs during the $\sphiswap$ portion of the gate, see Fig.~\ref{fig:sqrtiswap}(d). Closed-system simulations of this pulse sequence yield a gate fidelity of $F=0.9996$ for achieving a $\siswap$ gate, calculated using Eq.~\eqref{eq:F}, taking $d=4$ and $U_{\mathrm{T}}=\siswap$. Infidelities at the $10^{-4}$ level are likely due to residual effects from the higher-lying states that cause unwanted transitions in the computational subspace, see Appendix~\ref{appendix:tdH} for details.

To include the detrimental effects of decoherence on gate fidelities, we numerically solve the Lindblad master equation 
\begin{align}
	\frac{d\rho(t)}{dt}&=-i[H(t), \rho(t)] \\ \nonumber &\quad+\sum_{\mu=a,b}[\Gamma_{1}^{\mu}\mathcal{D}(L_{1}^{\mu})\rho(t) +\Gamma_{\phi}^{\mu}\mathcal{D}(L_{\phi}^{\mu})\rho(t)],
\end{align}
where $\rho(t)$ is the system density matrix and
\begin{align*}
	\mathcal{D}(L)\rho = L\rho L^{\dagger}-\frac{1}{2}\{L^{\dagger}L,\rho\},
\end{align*}
is the standard form of the dissipator. The relevant jump operators $L$ are here:
\begin{align*}
	L_{1}^{\mu}=\bar{\sigma}_{-}^{\mu},\quad L_{\phi}^{\mu}=\bar{\sigma}_{z}^{\mu}.
\end{align*}
We neglect decoherence processes due to higher-lying states, noting that their occupation remains minimal throughout the duration of the pulse. We consider two sets of estimates for decoherence rates, one conservative, $\Gamma_{\phi}=1/80\, \mu \text{s}, \Gamma_{1}=1/300\, \mu \text{s}$,
and one optimistic, $\Gamma_{\phi}=1/4000\, \mu \text{s}, \Gamma_{1}=1/1000\, \mu \text{s}$, both consistent with recent experiments \cite{Zhang2020, Somoroff2021}. At the conclusion of the gate, we project onto the computational subspace and perform numerical quantum process tomography \cite{qutip1, *qutip2, nesterov2021, nielsenchuang2010} to obtain the process matrix $\chi$. The open-system gate fidelity is calculated using the formula \cite{Pedersen2007, Chow2009, Nielsen2002, *Horodecki1999}
\begin{align}
F=\frac{d\Tr(\chi_{\mathrm{T}}\chi)+\Tr(\chi)}{d+1},
\end{align}
where $d$ is the dimension of the relevant subspace and $\chi_{\mathrm{T}}$ is the target process matrix. We obtain open-system gate fidelities of $F=0.997, F=0.9994$ for the two cases of conservative and optimistic decoherence rate estimates, respectively. 

\section{Discussion and Conclusion}
\label{sec:conc}

In this work, we have proposed a galvanic-coupling scheme for fluxonia in which $\textit{XX}$ coupling can be switched on and off while maintaining the qubits at their respective sweet spots. Motivated by record coherence times achieved with heavy-fluxonium qubits, we have concentrated on operating at frequencies below $\sim$200 MHz \cite{Zhang2020, Somoroff2021}. The magnitude of drive strengths required in this case invalidates RWA and makes it more natural to perform gates with reference to the lab frame. 
We have presented a protocol involving flux biasing and strong flux modulation that achieves a fast and high-fidelity $\sphiswap$ gate. To transform this into the more familiar $\siswap$ gate, we introduce variable-time identity gates.
These gates when combined with $\textit{Z}$ rotations help us realize a $\siswap$ gate with fidelity $F>0.999$. 
Fidelities are limited by incoherent errors as well as unwanted transitions in the computational subspace mediated by higher-lying states.

A crucial open question that warrants future research is
how to achieve scalability in this fluxonium-based architecture.
One may envision extending the device into a 1D array of qubits and couplers. Generalization to 2D arrays with increased qubit connectivity will require additional modifications, and will be useful for steps towards e.g. error-correcting surface codes \cite{Fowler2012}. 

\section{Acknowledgments}

We thank Brian Baker, Ziwen Huang and Yuan-Chi Yang for helpful discussions. D.~K.~W. acknowledges support from the Army Research Office (ARO) through a QuaCGR Fellowship. This research was funded by the ARO under Grant No.\ W911NF-19-10016. This work relied on multiple open-source software packages, including matplotlib \cite{matplotlib}, numpy \cite{numpy}, QuTiP \cite{qutip1, *qutip2}, scipy \cite{scipy}, and scqubits \cite{scqubits}.

\appendix

\section{Full-circuit analysis}
\label{appendix:fullcircuit}

\begin{figure}
    \centering
    \includegraphics[width=\columnwidth]{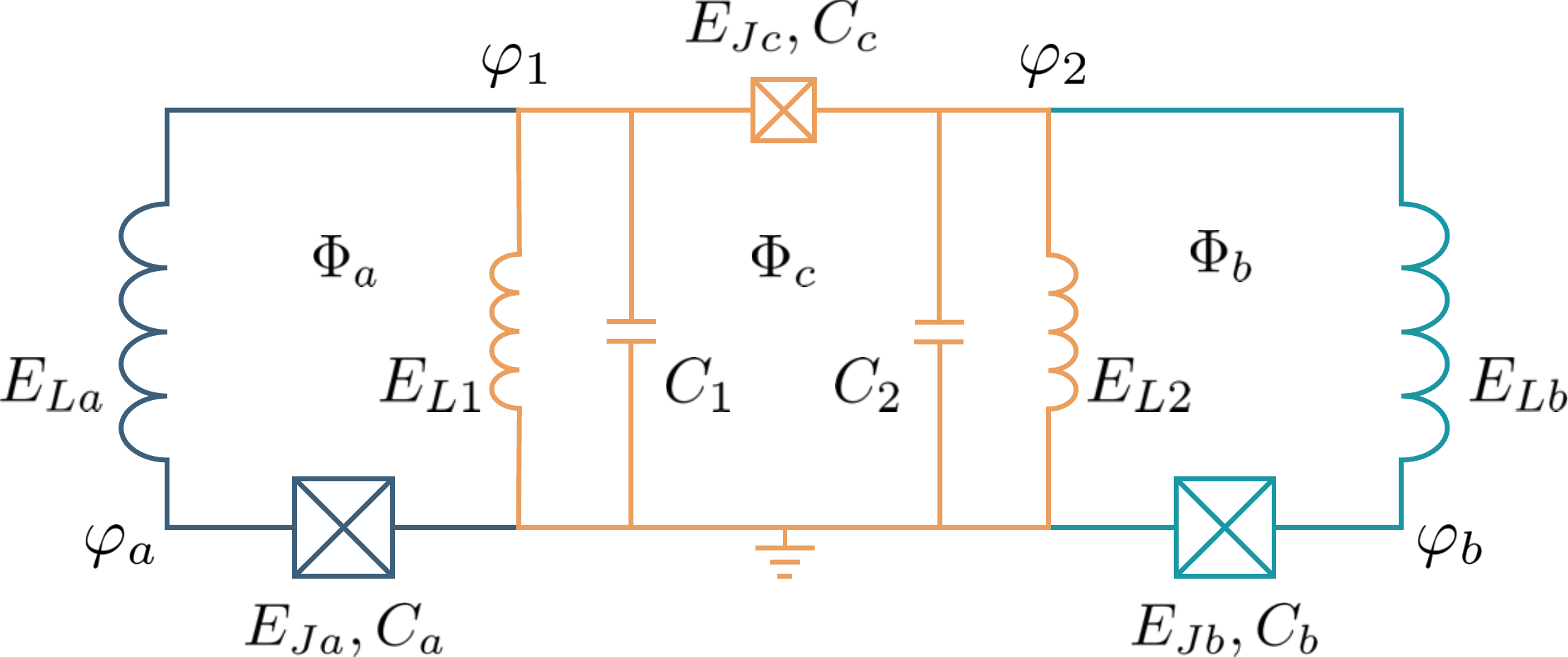}
    \caption{\label{fig:full_circuit} Schematic of the full circuit, accounting for parameter disorder.}
\end{figure}
In this appendix, we construct the Lagrangian and Hamiltonian of the full circuit shown in Fig.~\ref{fig:full_circuit}, allowing for disorder in circuit parameters.
We follow the method of Vool and Devoret \cite{Vool2017} to construct the circuit Lagrangian, yielding
\begin{align*}
\nonumber 
&\mathcal{L} =\frac{\Phi_0^2}{2(2\pi)^2}\left(\sum_{\mu=a,b}C_{\mu}\dot{\varphi}_{\mu}^2+\sum_{i=1,2}C_{i}\dot{\varphi}_{i}^2+C_{c}[\dot{\varphi}_{1}-\dot{\varphi}_{2}]^2\right) \\ 
&-\frac{1}{2}E_{La}(\varphi_{a}-\varphi_{1})^{2}-\frac{1}{2}E_{Lb}(\varphi_{b}-\varphi_{2})^{2} -\frac{1}{2}\sum_{i=1,2}E_{Li}\varphi_{i}^{2} \\ \nonumber 
    &+ \sum_{\mu=a,b}E_{J\mu}\cos(\varphi_{\mu}+\phi_{\mu}) + E_{Jc}\cos(\varphi_{1}-\varphi_{2} + \phi_{c}),
\end{align*}
with node variables and circuit parameters as shown in Fig.~\ref{fig:full_circuit}. 
We consider the case of small deviations from otherwise pairwise equivalent qubit inductors $E_{La},E_{Lb}$, coupler inductors $E_{L1},E_{L2}$ and stray capacitances $C_{1},C_{2}$. In the absence of parameter disorder, the coupler modes $\theta_{\pm}=\varphi_{1}\pm\varphi_{2}$ decouple, simplifying the analysis (though we show below that small parameter disorder is not expected to significantly impact device performance). Using these variables the Lagrangian becomes
\begin{align}
    \mathcal{L} &=\frac{1}{2}\left(\frac{\Phi_0}{2\pi}\right)^2\Bigg(\sum_{\mu=a,b}C_{\mu}\dot{\varphi}_{\mu}^2+\frac{1}{2}C\dot{\theta}_{+}^2 \\ \nonumber &\quad+[C_{c}+\frac{1}{2}C]\dot{\theta}_{-}^{2}+\frac{1}{2}C\,\mathrm{d}C\,\dot{\theta}_{+}\dot{\theta}_{-}\Bigg) \\ \nonumber 
    &\quad+\sum_{\mu=a,b}(E_{J\mu}\cos[\varphi_{\mu}+\phi_{\mu}]-\frac{1}{2}E_{L\mu}\varphi_{\mu}^{2}) \\ \nonumber 
    &\quad-\sum_{i=\pm}\frac{1}{2}E_{Lc}\theta_{i}^{2} + E_{Jc}\cos(\theta_{-} + \phi_{c}) \\ \nonumber
    &\quad-\frac{1}{4}(E_{L}\mathrm{d}E_{L}+E_{L}'\mathrm{d}E_{L}')\theta_{+}\theta_{-} \\ \nonumber
    &\quad+\frac{1}{2}E_{La}\varphi_{a}(\theta_{+}+\theta_{-})+\frac{1}{2}E_{Lb}\varphi_{b}(\theta_{+}-\theta_{-}) ,
\end{align}
where $E_{Lc}=\frac{1}{2}(E_{L}+E_{L}')$ and we have introduced notation for the average and relative deviation of the qubit inductors
$E_{L}=\frac{1}{2}(E_{La}+E_{Lb}),
\mathrm{d}E_{L} = \frac{E_{La}-E_{Lb}}{E_{L}}$,
the coupler inductors
$E_{L}'=\frac{1}{2}(E_{L1}+E_{L2}),
\mathrm{d}E_{L}' = \frac{E_{L1}-E_{L2}}{E_{L}'}$,
and the stray capacitances 
$C=\frac{1}{2}(C_{1}+C_{2}),
\mathrm{d}C = \frac{C_{1}-C_{2}}{C}$. We perform the Legendre transform to obtain the Hamiltonian and promote our variables to non-commuting operators obeying the commutation relations $[\varphi_{\mu},n_{\mu}]=i$ for $\mu=a,b$ and $[\theta_{j},n_{j}]=i$ for $j=\pm$. The Hamiltonian is $H=H_{0}+V+H_{\mathrm{dis}}$, where 
\begin{align}
\nonumber 
H_{0} &=\sum_{\mu=a,b}[4E_{C\mu}n_{\mu}^2+\frac{1}{2}E_{L\mu}\varphi_{\mu}^2 -E_{J\mu}\cos(\varphi_{\mu}+\pi)]\\ 
&\quad+4E_{C-}n_{-}^2+\frac{1}{2}E_{Lc}\theta_{-}^2-E_{Jc}\cos(\theta_{-}+\phi_{c}) \\ \nonumber 
&\quad+4E_{C+}n_{+}^2+\frac{1}{2}E_{Lc}\theta_{+}^2, \\
\label{eq:appV}
V &=-\sum_{\mu=a,b} \frac{E_{L\mu}}{2}\varphi_{\mu}[\theta_{+} + (-1)^{\mu}\theta_{-}]   \\ \nonumber
&\quad+\sum_{\mu=a,b}\frac{E_{L\mu}}{2}{\delta\phi^{\mu}}[-2\varphi_{\mu}+\theta_{+}+(-1)^{\mu}\theta_{-}], \\
\label{eq:Hdis}
H_{\mathrm{dis}} &=\frac{1}{4}(E_{L}\mathrm{d}E_{L}+E_{L}'\mathrm{d}E_{L}')\theta_{+}\theta_{-}
\\ \nonumber 
&\quad -4E_{C-}\mathrm{d}Cn_{+}n_{-} + \mathcal{O}(\mathrm{d}C^2).
\end{align}
Charging energy definitions introduced above are given in the main text and $H_{0}, V$ and $H_{\mathrm{dis}}$ refer to the bare, coupling and disorder Hamiltonians, respectively. We have neglected higher-order disorder contributions proportional to $\mathrm{d}C^2$ on the assumption that disorder is small. 
Additionally, we isolated the qubit flux shift away from the sweet spot $\delta\phi_{\mu}=\phi_{\mu}-\pi$ and performed the variable transformation $\varphi_{\mu}\rightarrow\varphi_{\mu}-\delta\phi_{\mu}$. 

\section{Schrieffer-Wolff transformation}
\label{appendix:SWT}

In this appendix, we derive the second-order effective Hamiltonian describing the qubit-qubit interaction that is dispersively mediated by the tunable coupler. First, we consider the symmetric case $H_{\text{dis}}=0$. Later, we relax this assumption and allow for parameter disorder. 

\subsection{Effective Hamiltonian without parameter disorder}
We separate the Hilbert space into a low- and high-energy subspace defined by the respective projectors
\begin{align}
P &=\sum_{\ell,m\in\{0,1\}}|\ell_{a},m_{b},0_{-},0_{+}\rangle\langle \ell_{a},m_{b},0_{-},0_{+}|, \\
Q &= \openone-P.
\end{align}
We have defined the states $|\ell_{a},m_{b},n_{-},p_{+}\rangle$ that are eigenstates of the bare Hamiltonian $H_{0}$ with eigenenergies $E_{\ell}^{a}+E_{m}^{b}+E_{n}^{-}+p\omega_{+}$.
The perturbation $V$ couples states within the same subspace, as well as states in separate subspaces. We utilize a Schrieffer-Wolff transformation \cite{Blais2021, SW, *cohentannoudji, *Winkler2003} $e^{-S}$ with anti-hermitian generator $S$ to decouple the low- and high-energy subspaces order-by-order. 
To carry out the transformation, the effective Hamiltonian $H_{\mathrm{eff}}=Pe^{S}He^{-S}P$ and generator $S$ are expanded as
\begin{align}
H_{\mathrm{eff}} &=H^{(0)} +  H^{(1)} + H^{(2)} + \cdots, \\
S &=  S^{(1)} + S^{(2)} + \cdots .
\end{align} 
We then collect terms of the same order and enforce both that the effective Hamiltonian is block diagonal and that the generator is block off diagonal. 

The zeroth- and first-order contributions to the effective Hamiltonian in the computational subspace (neglecting constant terms) are found by applying the  projector $P$ onto $H_{0}$ and $V$ respectively \cite{Blais2021, SW, *cohentannoudji, *Winkler2003}
\begin{align}
\label{eq:appH0}
H^{(0)} &= PH_{0}P =-\sum_{\mu=a,b}\frac{\omega_{\mu}}{2}\sigma_{z}^{\mu} \\
\label{eq:appH1}
H^{(1)} &= PVP = -\sum_{\mu=a,b}\Omega_{\mu}\sigma_{x}^{\mu} 
\end{align}
where $\omega_{\mu} =E_{1}^{\mu}-E_{0}^{\mu}$ and
\begin{align}
\label{eq:app_Omega}
\Omega_{\mu}=E_{L}\langle 0_{\mu}|\varphi_{\mu}|1_{\mu}\rangle\left[\delta\phi_{\mu}+(-1)^{\mu}\frac{\langle 0_{\mu}|\theta_{-}|0_{\mu}\rangle}{2}\right].
\end{align}
The Pauli matrices are defined as e.g. $\sigma_{x}^{a}=\sum_{m=0,1}|0_{a},m_{b},0_{-},0_{+}\rangle\langle1_{a},m_{b},0_{-},0_{+}|+\mathrm{H.c.}$ 

Calculation of the second-order contribution $H^{(2)}$ to the effective Hamiltonian is facilitated by the first-order generator $S^{(1)}$. The expression for the matrix elements of $S^{(1)}$ is well known \cite{Blais2021} and we obtain
\begin{align}
\label{eq:S1}
S^{(1)} =\sum_{\substack{\mu=a,b\\j=0,1}}\Bigg(&\sideset{}{'}\sum_{j',n} (-1)^{\mu+1}\epsilon^{\mu,(1)}_{jj',n}
|j_{\mu},0_{-}\rangle\langle j_{\mu}', n_{-}|
\\ \nonumber 
- &\sum_{j'} \eta^{\mu,(1)}_{jj'}
|j_{\mu}, 0_{+}\rangle\langle j_{\mu}', 1_{+}|\Bigg) - \mathrm{H.c.},
\end{align}
defining the small parameters
\begin{align}
\label{eq:appeps}
\epsilon^{\mu,(1)}_{jj',n} &= \frac{g^{\mu}_{jj',0n}}{
E_{jj'}^{\mu}-E_{n0}^{-}}, \quad
\eta^{\mu,(1)}_{jj'} = \frac{G^{\mu}_{jj'}}{E_{jj'}^{\mu}-\omega_{+}},
\end{align}
where $E_{jk}^{\mu}=E_{j}^{\mu}-E_{k}^{\mu}, \mu=a,b,-,$ and
\begin{align}
g_{jj',0n}^{\mu} &=\frac{E_{L}}{2}\langle 0_{-}|\theta_{-}|n_{-}\rangle\langle j_{\mu}|\varphi_{\mu}|j'_{\mu}\rangle, \\ 
G_{jj'}^{\mu} &= \frac{E_{L}}{2}\left(\frac{2E_{C+}}{E_{Lc}}\right)^{1/4}\langle j_{\mu}|\varphi_{\mu}|j'_{\mu}\rangle.
\end{align}
We have introduced annihilation and creation operators $a_{+}, a_{+}^{\dagger}$ for the coupler $\theta_{+}$ mode via $\theta_{+}=\frac{\ell_{\mathrm{osc}}}{\sqrt{2}}(a_{+}+a_{+}^{\dagger})$ and $\ell_{\mathrm{osc}}=(8E_{C+}/E_{Lc})^{1/4}$ is the oscillator length. The primed sum in Eq.~\eqref{eq:S1} indicates that $n$ is allowed to be zero if $j'\geq2$, acknowledging that the perturbation $V$ can couple computational states to higher-lying states of the qubit fluxonia without exciting the coupler $\theta_{-}$ mode. We have neglected contributions proportional to $\delta\phi_{\mu}$ in Eq.~\eqref{eq:S1} as they are comparatively small and can be neglected.

Using the first-order generator, we can compute the second-order effective Hamiltonian in the low-energy subspace via the formula $H^{(2)}=\frac{1}{2}P[S^{(1)},V]P$ \cite{Blais2021, SW, *cohentannoudji, *Winkler2003}, yielding
\begin{align}
\label{eq:H2}
H^{(2)} &=-\sum_{\mu=a,b}\frac{\chi_{\mu}}{2}\sigma_{z}^{\mu}+J\sigma_{x}^{a}\sigma_{x}^{b},
\end{align}
where we have defined $\chi_{\mu}=\chi_{1}^{\mu}-\chi_{0}^{\mu}$ and neglected global energy shifts. 
The qubit-frequency renormalization coefficients are
\begin{align}
\label{eq:ximu}
\chi^{\mu}_{j}&=-\sum_{j'}\left(\sideset{}{'}\sum_{n}\frac{|g_{jj',n0}^{\mu}|^2}{\delta_{jj',n}^{\mu}}
+\frac{|G_{jj'}^{\mu}|^2}{\Delta_{jj'}^{\mu}}\right),
\end{align}
defining the energy denominators $\delta_{jj',n}^{\mu}=E_{n0}^{-}-E_{jj'}^{\mu}$ and $\Delta_{jj'}^{\mu}=\omega_{+}-E_{jj'}^{\mu}$.
The two-qubit interaction strength is
\begin{align}
\label{eq:Jmaintext}
\nonumber
J &= \sum_{n\geq1}\frac{g_{01,0n}^{a}g_{01,0n}^{b}}{2}\left(\frac{1}{\delta_{01, n}^{a}} + \frac{1}{\delta_{10, n}^{a}}+\frac{1}{\delta_{01, n}^{b}} + \frac{1}{\delta_{10, n}^{b}} \right) \\  &\quad-\frac{G_{01}^{a}G_{01}^{b}}{2}\left(\frac{1}{\Delta_{01}^{a}} + \frac{1}{\Delta_{10}^{a}}+\frac{1}{\Delta_{01}^{b}} + \frac{1}{\Delta_{10}^{b}} \right) \\ \nonumber 
&=  J_{-} - J_{+},
\end{align}
implicitly defining $J_{\pm}$.
Thus, the effective Hamiltonian in the computational subspace up to second order is 
\begin{align}
\label{eq:appHeff}
H_{\mathrm{eff}}=-\sum_{\mu=a,b}\frac{\omega_{\mu}'}{2}\sigma_{z}^{\mu}+J\sigma_{x}^{a}\sigma_{x}^{b} - \sum_{\mu=a,b}\Omega_{\mu}\sigma_{x}^{\mu},
\end{align}
where $\omega_{\mu}'=\omega_{\mu}+\chi_{\mu}$. 

\subsection{Effective Hamiltonian in the presence of disorder}

We now consider how disorder in circuit parameters modifies the form of the effective Hamiltonian Eq.~\eqref{eq:appHeff}. This disorder could arise for example from fabrication imperfections. We show below that up to second order, inductive disorder merely results in a modification to Eq.~\eqref{eq:app_Omega}, while capacitive disorder does not contribute.

From Eq.~\eqref{eq:Hdis} we see that inductive asymmetry adds a disorder term to the Hamiltonian 
\begin{align}
H_{\mathrm{ind}}= \frac{1}{4}(E_{L}dE_{L}+E_{L}'dE_{L}')\theta_{+}\theta_{-}.
\end{align}
If we assume that the relative deviations are small compared with unity, it is justified to add this term to $V$ and treat it perturbatively. Observe that on the one hand, for virtual transitions mediated by this term, the excitation number of either qubit cannot change. On the other hand, the excitation number of the coupler $\theta_{+}$ mode must change. Thus, the first-order contributions vanish, and the only second-order terms that contribute beyond a global energy shift are
\begin{align}
\label{eq:appH2dis}
H^{(2)}_{\mathrm{ind}} =
-\frac{1}{2}\sum_{\mu=a,b}g_{\mathrm{ind}}(\eta^{\mu}_{01}+\eta^{\mu}_{10})\sigma_{x}^{\mu},
\end{align}
where we have defined
\begin{align}
\label{eq:appgdis}
g_{\mathrm{ind}} = \frac{\ell_{\mathrm{osc}}}{4\sqrt{2}}(E_{L}dE_{L}+E_{L}'dE_{L}')\langle0_{-}|\theta_{-}|0_{-}\rangle.
\end{align}
Thus up to second order, disorder in the inductors serves only to modify the expressions for the coefficients $\Omega_{\mu}$. As discussed in the main text, this amounts to a shift in the sweet spot location of each qubit and can be canceled by a corresponding shift of the static qubit fluxes. Thus, small disorder in either the qubit inductors or the coupler inductors does not adversely affect device performance.

We now turn our attention to capacitive disorder $C_{1}\neq C_{2}$ (disorder in the qubit capacitances poses no issue, as the qubits remain decoupled from all other degrees of freedom in the kinetic part of the Hamiltonian). In this case, we proceed as before and treat perturbatively the capacitive disorder term
\begin{align}
\label{eq:Hcapdiss}
H_{\mathrm{cap}}=-4E_{C-}dCn_{+}n_{-}.
\end{align}
Consider the relation between phase and charge matrix elements in fluxonium \cite{Nesterov2018}
\begin{align}
\langle j_{-} | n_{-} | k_{-}\rangle = i \frac{E_{jk}^{-}}{8E_{C-}}\langle j_{-} | \theta_{-}| k_{-}\rangle,
\end{align}
and observe that the charge matrix element vanishes if $j=k$. Thus, any virtual transition mediated by the perturbation \eqref{eq:Hcapdiss} must excite both the coupler $\theta_{-}$ mode and the coupler $\theta_{+}$ mode and thus does not contribute at second order beyond a global energy shift.

\section{Drive operators}
\label{appendix:tdH}

In this appendix, we calculate the matrix elements and consider the effects of the relevant drive operators activated by time-dependent flux drives. Allocating the time-dependent flux in the same way as for static flux generally introduces terms proportional to the time derivative of the external flux \cite{You2019}. Imposing the constraint that these terms should not appear implies a specific grouping of the flux in the full Hamiltonian $H$. For our parameters, we find to a good approximation that the ac qubit fluxes are allocated to their respective inductors, and the ac coupler flux is spread across all four inductors. 
We first decompose the external fluxes into static $\bar{\phi}_{\mu}$ and time-dependent $\delta\phi_{\mu}(t)$ components, where $\bar{\phi}_{\mu}$ are the dc flux values at the off position.
The ac qubit fluxes are already properly allocated, while the appropriate grouping of the coupler flux is obtained via $\theta_{-}\rightarrow\theta_{-}-\delta\phi_{c}(t)$.
The full time-dependent Hamiltonian is thus $H(t)=H_{0}+V+h_{a}\delta\phi_{a}(t)+h_{b}\delta\phi_{b}(t)+h_{c}\delta\phi_{c}(t)$, where
\begin{align}
\label{eq:Ha}
h_{a} &=\frac{E_{L}}{2}(-2\varphi_{a}+\theta_{+}+\theta_{-}), \\
\label{eq:Hb}
h_{b} &=\frac{E_{L}}{2}(-2\varphi_{b}+\theta_{+}-\theta_{-}), \\
\label{eq:Hc}
h_{c} &=\left(\frac{E_{L}}{2}\varphi_{a}-\frac{E_{L}}{2}\varphi_{b}-E_{Lc}\theta_{-}\right).
\end{align}
Matrix elements of the operators $h_{\mu}$ with respect to eigenstates of the static Hamiltonian $H_{\mathrm{st}}=H_{0}+V$ determine the time evolution, once the time-dependent drives $\delta\phi_{\mu}(t)$ are specified. The Schrieffer-Wolff transformation allows us to define new basis states that are approximate eigenstates of $H_{\mathrm{st}}$ and thus perturbatively calculate these matrix elements.

The leading-order contributions to select matrix elements of the drive operators $h_{\mu}$ occur at second order. Thus, to include all relevant corrections to the wave functions that contribute to these matrix elements, we calculate the second-order generator $S^{(2)}$ associated with the Schrieffer-Wolff transformation discussed in Appendix~\ref{appendix:SWT}. To simplify the calculation we ignore all contributions from the coupler $\theta_{+}$ mode due to the inequality $|\eta|<|\epsilon|$ in parameter regimes of interest, yielding \cite{Blais2021, Winkler2003}
\begin{align}
&S^{(2)} =
\sum_{\substack{\mu=a,b\\j=0,1}}\sideset{}{'}\sum_{j',n}\epsilon^{\mu,(2)}_{jj',n}|j_{\mu},0_{-}\rangle\langle j_{\mu}',n_{-}| \\ \nonumber 
&+\sum_{j,k=0,1} \sideset{}{'}\sum_{j',k',n}\epsilon^{ab,(2)}_{jj',kk',n}|j_{a},k_{b},0_{-}\rangle\langle j_{a}',k_{b}',n_{-}| -\mathrm{H. c.} 
\end{align}
where we have defined
\begin{align}
\nonumber 
\epsilon^{\mu,(2)}_{jj',n} &=-\sum_{j''=0,1}\frac{g^{\mu}_{jj'',00}g^{\mu}_{j''j',0n}}{(E_{jj'}^{\mu}-E_{n0}^{-})(E_{j''j'}^{\mu}-E_{n0}^{-})}
\\ \nonumber 
&\quad+\sideset{}{'}\sum_{j'',n'}\frac{g^{\mu}_{jj'',0n'}g^{\mu}_{j''j',n'n}}{(E_{jj''}^{\mu}-E_{n'0}^{-})(E_{jj'}^{\mu}-E_{n0}^{-})}, \\ \nonumber 
\epsilon^{ab,(2)}_{jj',kk',n} &=
\sum_{\substack{\mu,\nu=a,b\\\mu\neq\nu}}\Bigg[\frac{g^{\mu}_{jj',00}g^{\nu}_{kk',0n}}{(E_{jj'}^{\mu}+E_{kk'}^{\nu}-E_{n0}^{-})(E_{kk'}^{\nu}-E_{n0}^{-})} \\ \nonumber 
&\quad-\sideset{}{'}\sum_{n'}\frac{g^{\mu}_{jj',0n'}g^{\nu}_{kk',n'n}}{(E_{jj'}^{\mu}-E_{n'0}^{-})(E_{jj'}^{\mu}+E_{kk'}^{\nu}-E_{n0}^{-})}\Bigg].
\end{align}

At the off position, the effective Hamiltonian $H_{\mathrm{eff}}=-\frac{\omega_{a}'}{2}\sigma_{z}^{a}-\frac{\omega_{b}'}{2}\sigma_{z}^{b}$ (ignoring third-order contributions to the effective Hamiltonian) is diagonal in the basis of the bare computational states $|\ell_{a},m_{b},0,0\rangle, \ell,m\in\{0,1\}$. 
Assuming the qubit frequencies are not on resonance $\omega_{a}'\neq\omega_{b}'$, the dressed eigenstates are
\begin{align}
\label{eq:dressed}
&|\overline{\ell_{a},m_{b},0_{-},0_{+}}\rangle = e^{-S}|\ell_{a},m_{b},0_{-},0_{+}\rangle \\ \nonumber 
&= \left[\openone-S^{(1)}-S^{(2)}+\frac{1}{2}(S^{(1)})^2\right] |\ell_{a},m_{b},0_{-},0_{+}\rangle,
\end{align}
up to second order. With the dressed eigenstates now written in terms of bare states, we may compute matrix elements of the operators $h_{\mu}$ associated with the ac flux drives. 

\subsection{Qubit-flux drive operators}

Experimentally, the amplitude of an ac flux drive will typically be no larger than $\delta\Phi_{\mu}\leq 0.1 \Phi_{0}$ \cite{Zhang2020}. In this case, we have checked that transitions to higher-lying states mediated by the drive operators $h_{a}, h_{b}$ are suppressed. Thus, we need only consider matrix elements of these operators in the computational subspace. Using the expression \eqref{eq:dressed} for the dressed states given in terms of the bare states, we find
\begin{widetext}
\begin{align}
\label{eq:A1}
\langle\overline{\ell m}|h_{a}|\overline{\ell m}\rangle/E_{L} &=2\sum_{\ell'\geq2}\langle\ell_{a}|\varphi_{a}|\ell'_{a}\rangle\epsilon^{a,(1)}_{\ell\ell',0}+\frac{1}{2}\langle0_{-}|\theta_{-}|0_{-}\rangle, \\
\label{eq:A2}
\langle\overline{\ell m}|h_{a}|\overline{\ell+1 m}\rangle/E_{L} &=
-\langle0_{a}|\varphi_{a}|1_{a}\rangle -\frac{1}{2}\sum_{n\geq1}\langle 0_{-}|\theta_{-}|n_{-}\rangle (\epsilon^{a,(1)}_{01,n}+\epsilon^{a,(1)}_{10,n}) - \frac{1}{2}\frac{\ell_{\mathrm{osc}}}{\sqrt{2}}(\eta^{a,(1)}_{01}+\eta^{a,(1)}_{10}),\\
\label{eq:A3}
\langle\overline{\ell m}|h_{a}|\overline{\ell m+1}\rangle/E_{L} &=
-\frac{1}{2}\frac{\ell_{\mathrm{osc}}}{\sqrt{2}}(\eta^{b,(1)}_{01}+\eta^{b,(1)}_{10}) +\frac{1}{2}\sum_{n\geq1}\langle 0_{-}|\theta_{-}|n_{-}\rangle(\epsilon^{b,(1)}_{01,n}+\epsilon^{b,(1)}_{10,n})], \\
\label{eq:A4}
\langle\overline{\ell m}|h_{a}|\overline{\ell+1 m+1}\rangle/(E_{L}/2) &=
-\sum_{n,n'\geq1}(\epsilon^{a,(1)}_{\ell\ell+1,n}\epsilon^{b,(1)}_{m+1m,n'} +\epsilon^{a,(1)}_{\ell+1\ell,n}\epsilon^{b,(1)}_{mm+1,n'})\langle n_{-}|\theta_{-}|n'_{-}\rangle \\ \nonumber 
+\sum_{n\geq1}(\epsilon^{ab,(2)}_{\ell\ell+1,mm+1,n} &+\epsilon^{ab,(2)}_{\ell+1\ell,m+1m,n})\langle0_{-}|\theta_{-}|n_{-}\rangle +\sum_{n\geq1}(\epsilon^{a,(1)}_{\ell\ell+1,n}\epsilon^{b,(1)}_{m+1m,n}+\epsilon^{a,(1)}_{\ell+1\ell,n}\epsilon^{b,(1)}_{mm+1,n})\langle0_{-}|\theta_{-}|0_{-}\rangle,
\end{align}
\end{widetext}
where we have introduced the shorthand $|\overline{\ell,m}\rangle\equiv|\overline{\ell_{a},m_{b},0_{-},0_{+}}\rangle$ for states in the computational subspace, and the labels are understood modulo 2. 
In Eqs.~\eqref{eq:A1}-\eqref{eq:A3}, second-order contributions are small and can be neglected, while in Eq.~\eqref{eq:A4} the leading-order contributions are at second-order. These analytical approximations indicate that in the computational subspace and at the off position, the operator $h_{a}$ simplifies dramatically to leading order to $\Omega_{\mathrm{ac}}^{a}\bar{\sigma}_{x}^{a}$. For example, the matrix element \eqref{eq:A3} vanishes at the off position by definition, see Eq.~\eqref{eq:Jmaintext}. Further explicit verification of the form of $h_{a}$ is tedious and will be omitted.
We find excellent agreement between the semi-analytic formulas and exact results: for the parameters considered here, we obtain $|\Omega_{\mathrm{ac}}^{a}|/h=558, 561\; $ MHz using Eqs.~\eqref{eq:A1}-\eqref{eq:A4} and numerics, respectively. The coefficients associated with all other operators (aside from the irrelevant identity) in the decomposition of $h_{a}$ are of the order of 2 $h\cdot$MHz or smaller in absolute value, as calculated both from the semi-analytic formulas and exact results. 

\subsection{Coupler-flux drive operator}

The operator $h_{c}$ activated by coupler-flux modulation induces both wanted and unwanted transitions in the computational subspace. The latter proceed through virtual excitations of higher-lying states. 
We analyze both types of transitions in the following.

\subsubsection{Computational-subspace matrix elements}

The matrix elements of $h_{c}$ governing the wanted transitions can be obtained within second-order perturbation theory using Eq.~\eqref{eq:dressed}
\begin{widetext}
\begin{align}
\label{eq:CII}
\langle\overline{\ell m}|h_{c}|\overline{\ell m}\rangle &= -E_{Lc}\langle0_{-}|\theta_{-}|0_{-}\rangle -E_{L}\sum_{\ell'\geq2}\langle\ell_{a}|\varphi_{a}|\ell'_{a}\rangle\epsilon^{a,(1)}_{\ell\ell',0} -E_{L}\sum_{m'\geq2}\langle m_{b}|\varphi_{b}|m'_{b}\rangle\epsilon^{b,(1)}_{mm',0},
 \\
\label{eq:CXI}
\langle\overline{\ell m}|h_{c}|\overline{\ell+1 m}\rangle &=  \frac{E_{L}}{2}\langle0_{a}|\varphi_{a}|1_{a}\rangle +E_{Lc}\sum_{n\geq1}\langle0_{-}|\theta_{-}|n_{-}\rangle (\epsilon^{a,(1)}_{01,n}+\epsilon^{a,(1)}_{10,n}), \\
\label{eq:CIX}
\langle\overline{\ell m}|h_{c}|\overline{\ell m+1}\rangle &= -\frac{E_{L}}{2}\langle 0_{b}|\varphi_{b}| 1_{b}\rangle -E_{Lc}\sum_{n\geq1}\langle0_{-}|\theta_{-}|n_{-}\rangle(\epsilon^{b,(1)}_{01,n}+\epsilon^{b,(1)}_{10,n}), \\
\label{eq:CXX}
\langle\overline{\ell m}|h_{c}|\overline{\ell+1 m+1}\rangle 
&=-\frac{E_{Lc}}{E_{L}/2}\langle\overline{\ell m}|h_{a}|\overline{\ell+1 m+1}\rangle.
\end{align}
\end{widetext}
At the off position $h_{c}$ can be simplified to 
\begin{align}
h_{c}=J_{\mathrm{ac}}\bar{\sigma}_{x}^{a}\bar{\sigma}_{x}^{b},
\end{align}
where we have defined the ac $\textit{XX}$ coupling strength
\begin{align}
\label{eq:Jac}
J_{\mathrm{ac}}=\frac{1}{2}(\langle\overline{00}|h_{c}|\overline{11}\rangle+\langle\overline{01}|h_{c}|\overline{10}\rangle).
\end{align}
For our parameters, we obtain $|J_{\mathrm{ac}}|/h=14.3, 18.3\; $ MHz using the semi-analytic formulas Eqs.~\eqref{eq:CII}-\eqref{eq:CXX} and exact numerics, respectively.
We have checked that the semi-analytic results agree with exact numerics in the limit of large $E_{Lc}$, where the interaction becomes more dispersive. 

\subsubsection{Virtual transitions involving higher-lying states}

\begin{figure}
    \centering
    \includegraphics[width=0.9\columnwidth]{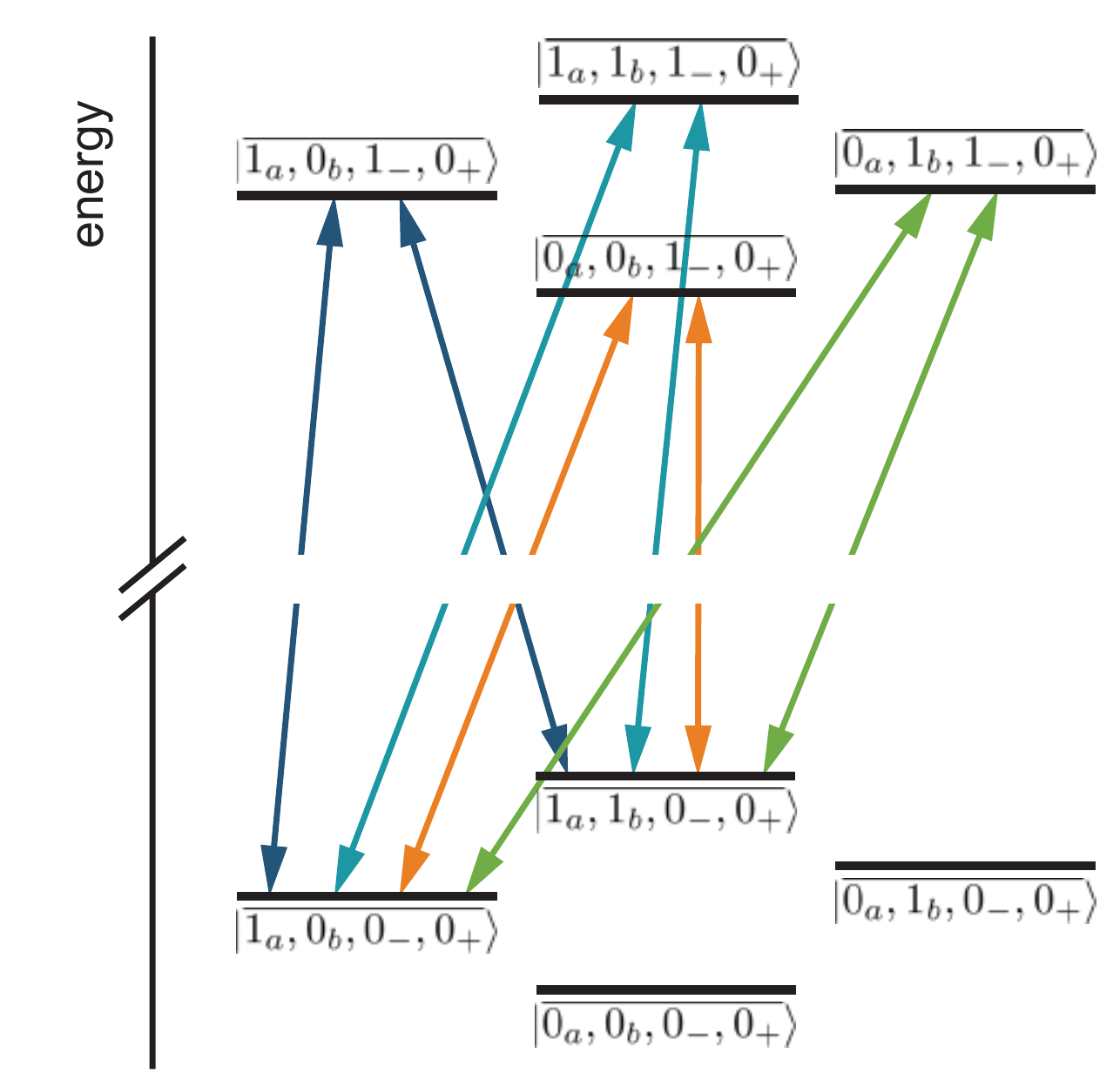}
    \caption{\label{fig:unwanted_trans} Schematic of unwanted transitions in the computational subspace. These transitions are due to virtual excitations of the higher-lying states $\oket{\ell_{a},m_{b},1_{-},0_{+}}, \ell,m\in\{0,1\}$ mediated by the drive operator $h_{c}$. As an example, we show the four perturbative paths contributing to the undesired transition $\oket{1_{a},0_{b},0_{-},0_{+}}\leftrightarrow\oket{1_{a},1_{b},0_{-},0_{+}}$. }
\end{figure}

The full analysis of time evolution when modulating the coupler flux [Sec.~\ref{sec:two_q_gate}] requires consideration of higher-lying states. These states outside the computational subspace, while largely remaining unoccupied, participate as virtual intermediate states in unwanted transitions. We estimate the amount of population transfer between the states $|i\rangle=\oket{\ell_{a},m_{b},0_{-},0_{+}}$ and $|f\rangle=\oket{\ell'_{a},m'_{b},0_{-},0_{+}}$ with $i\neq f$ at the conclusion of the gate $t=2\pi n/\omega_{d}$ 
using time-dependent perturbation theory up to second-order \cite{sakurai}
\begin{align}
\label{eq:badtransitions}
&T_{\ell m\rightarrow \ell' m'} = 
\Bigg|\frac{i\langle f | \delta\phi_{c}h_{c} | i \rangle \omega_{d}}{E_{fi}^{2}-\omega_{d}^2}(1-e^{2\pi i n E_{fi}/\omega_{d}})\\ \nonumber 
&+\sum_{v}
\frac{\omega_{d}^2\langle f|\delta\phi_{c}h_{c}|v\rangle\langle v|\delta\phi_{c}h_{c}|i\rangle}{E_{vi}^2-\omega_d^2} \\ \nonumber 
&\Bigg(\frac{[2E_{vi}+E_{fi}][1-e^{2\pi n i E_{fi}/\omega_{d}}]}{E_{fi}[E_{fi}^2-4\omega_{d}^2]}-\frac{1-e^{2\pi n i E_{fv}/\omega_{d}}}{E_{fv}^2-\omega_{d}^2}\Bigg)\Bigg|^2,
\end{align}
where the sum on $v$ is over virtual intermediate states $|v\rangle=\oket{\ell_{a}'',m_{b}'',p_{-},q_{+}}$ and we have defined $E_{fv}=E_{\overline{\ell'm'00}}-E_{\overline{\ell''m''pq}}$, etc.
The top line of Eq.~\eqref{eq:badtransitions} represents direct transitions between the states $|i\rangle$ and $|f\rangle$, occurring for nonzero $\langle f | h_{c} | i \rangle$ (e.g. $|i\rangle=\oket{0_{a},1_{b},0_{-},0_{+}}$ and $|f\rangle=\oket{1_{a},0_{b},0_{-},0_{+}}$). These are the wanted transitions discussed previously. The second and third lines of Eq.~\eqref{eq:badtransitions} are the second-order contributions and allow for unwanted transitions.
Based on numerical calculation of the matrix elements of $h_{c}$ between the computational states and higher-lying states, we find that the four states $|v\rangle=|\overline{\ell_{a},m_{b},1_{-},0_{+}}\rangle, \ell,m\in \{0,1\}$ with an excitation in the $\theta_{-}$ mode dominate the sum on $v$, see Fig.~\ref{fig:unwanted_trans}.
(Note that this virtual process is heavily suppressed in the context of qubit-flux drives, due to the comparatively small coefficient $E_{L}/2\ll E_{Lc}$ multiplying the operator $\theta_{-}$ in $h_{a}$).

These transitions mediated by the higher-lying states can significantly degrade gate fidelities. Indeed, attempting to implement the $\sphiswap$ gate with only a single drive period leads to poor fidelities $F<0.9$ due to the unwanted transitions. Slowing down the gate by utilizing two drive periods as in Sec.~\ref{sec:two_q_gate} mitigates this issue in large part (infidelities are on the order of $10^{-4}$) by reducing the required drive amplitude. 
It is an interesting avenue for further research to investigate means for overcoming this limitation on $\delta\phi_{c}$ to achieve faster gate times without sacrificing fidelity.

\bibliography{bib}
\end{document}